\documentclass{llncs}

\usepackage{url}
\usepackage{xspace}
\usepackage{wrapfig}
\usepackage{shared-deps}
\usepackage{hyperref}
\usepackage{cleveref}
\usepackage{longtable}
\usepackage{nicefrac}
\crefformat{equation}{(#2#1#3)}
\Crefformat{equation}{(#2#1#3)}

\hypersetup{
  breaklinks   = true,
  colorlinks   = true, %Colours links instead of ugly boxes
  urlcolor     = blue, %Colour for external hyperlinks
  linkcolor    = blue, %Colour of internal links
  citecolor    = red   %Colour of citations
}
\hypersetup{final} % Needed to generate hyperlinks even in draft mode (ERROR IN MAC)

\begin{document}

\title{Formal Analysis of Lending Pools in Decentralized Finance}

\author{
Massimo Bartoletti\inst{1}\orcidID{0000-0003-3796-9774}
\and James Chiang\inst{2}\orcidID{0000-0002-5126-9494}
\and Tommi Junttila\inst{3}%\orcidID{}
\and \\ Alberto Lluch~Lafuente\inst{2}\thanks{Corresponding author.} \orcidID{0000-0001-7405-0818} 
\and Massimiliano Mirelli\inst{2}\orcidID{0000-0001-9441-173X}
\and Andrea Vandin\inst{4,2}\orcidID{0000-0002-2606-7241}
}

\institute{
Universit\`a degli Studi di Cagliari, Cagliari, Italy\\
\email{bart@unica.it}
\and Technical University of Denmark, DTU Compute, Copenhagen, Denmark
\email{\{jchi,albl\}@dtu.dk}
\newline
\email{massimilianomirelli.mm@gmail.com} 
\and Aalto University, Espoo, Finland\\
\email{tommi.junttila@aalto.fi }
\and Sant'Anna School of Advanced Studies, Pisa, Italy\\
\email{andrea.vandin@santannapisa.it}
}

\maketitle

\begin{abstract}
Decentralised Finance (DeFi) applications constitute an entire financial ecosystem deployed on blockchains. Such applications are based on complex protocols and incentive mechanisms whose financial safety is hard to determine. Besides, their adoption is rapidly growing, hence imperilling an increasingly higher amount of assets. Therefore, accurate formalisation and verification of DeFi applications is essential to assess their safety.
We have developed a tool for the formal analysis of one of the most widespread DeFi applications: Lending Pools (LP). This was achieved by leveraging an existing formal model for LPs,  the Maude verification environment and the MultiVeStA statistical analyser. The tool supports several analyses including reachability analysis, LTL model checking and statistical model checking. In this paper we show how the tool can be used to analyse several parameters of LPs that are fundamental to assess and predict their behaviour. In particular, we use statistical analysis to search for threshold and reward parameters that minimize the risk of unrecoverable loans.
      
\end{abstract}

\section{Introduction}
\label{chap:intro}
% Motivation
% Why verification in this area is essential
% 0. loss of funds (check SoK: DeFi + SoK: LP)

% Mention research questions:
% 1. LP model in Maude (% 2. research area is lacking of general models for these solutions)
% 2. statistical analysis question (define/motivate optimality criterion)

Financial trading has recently shifted to virtual markets, platforms entirely regulated and controlled by novel protocols. 
\textit{Decentralised Finance} (DeFi) \cite{ethereum2021defi} applications are deployed on blockchains like Ethereum~\cite{ethereum2021defi,buterin2013white}, which offer distributed infrastructures to execute \textit{smart contracts} \cite{eth2021smartcontracts} without intermediaries. 
%
%\commb{Although the sentence is true (strictly speaking), I propose to de-emphasize Ethereum, since DeFi application are mainstream also in other blockchains}
%
DeFi has recently been employed by a growing community of users. 
As of April 2022, the growth of the capital locked by DeFi applications has increased almost $10$ times in the last two years: from approximately \$9.78bn, on 1 April 2020, to over \$83.51bn, on 1 April 2022 \cite{defipulse2021web}. 
Even assuming the security guarantees ensured by the underlying blockchain, DeFi smart contracts have several vulnerabilities latent in their design \cite{qin2020attacking,zhou2021just}.  
Given the considerable amount of funds daily exchanged on DeFi platforms \cite{aave2021markets,compound2021markets}, even minor design flaws could determine massive and intolerable losses \cite{compound2020price-oracle}.
Notwithstanding the increasing interest of several research groups in this area \cite{bigi2015validation,bai2018formal,abdellatif2018formal,tolmach2021formal,angeris2019analysis,gudgeon2020decentralized}, 
the complexity of DeFi protocols
%
%\commb{begin remove}and their recentness constantly \commb{end remove} 
%
yields new interesting research problems. Formal verification of these systems is crucial, in order to ensure their correctness and security. 

The verification tool proposed in this paper simulates and analyses \textit{Lending Pools} (LPs), one of the most popular DeFi applications, whose two main features  are lending and borrowing assets, to support various financial practices, including margin trading. Our verification tool is based on the  formal model of LPs proposed in~\cite{bartoletti2020sok}. Such model encompasses the behaviour of the most widespread LPs, namely Aave \cite{aave2020white} and Compound \cite{leshner2019compound}\footnote{At the time of writing, these are the first and third as per the amount of capital locked \cite{defipulse2021web}.}. 

We craft an operational specification of the LP model of \cite{bartoletti2020sok} in Maude~\cite{clavel2020maude}, 
a specification language which is particularly suitable for highly concurrent systems such as LPs. 
Additionally, Maude provides a very extensive environment for both simulating and verifying the properties of the specified models. Given the complexity of the modelled systems, the analyses techniques offered by the Maude environment are not sufficient. 
Specifically, since the system may evolve by following an infinite number of execution paths, the traditional model checking methods result in being either ineffective or unviable. Therefore, the Maude-based LP simulator has been extended to support statistical analyses. This has been achieved by integrating the simulator with the MultiVeStA statistical analyser proposed in  \cite{sebastio2013multivesta} and recently redesigned to focus on analyses of interest for of economical agent-based models~\cite{vandin2021automated}. The tool offers analysis techniques from the family of statistical model checking~\cite{agha2018survey}. 
These statistical analyses, despite producing less accurate results, allow to observe the quantitative behaviour of large instances of the model, offering statistically-reliable results. In the case of lending pools, this approach allows to estimate parameters of the model so to increase its safety. Specifically, an essential safety property of the model is that the value of non-repayable loans is low.

This paper is based on the work done in~\cite{Mirelli2021} and proposes a Maude-based LP simulator (\Cref{chap:agents}) capable of conducting several analyses of lending pools including LTL model checking and statistical analysis.  
The tool is open source and available at~\cite{mirelli2021maudelp}.  
Additionally, the study showcases the usage of the tool by answering a still non-investigated research question, aiming at an enhancement of the analysed platforms' safety. In particular, the statistical analysis presented in~\Cref{chap:tool} shows that a choice of the parameters used to instantiate the LP model reduces the amount of non-repayable loans.          
\section{Lending Pools and Price Models}
\label{chap:bg}

\subsubsection{Lending Pools.}
\textit{Lending Pools}~\cite{werner2021sok}
%\commb{begin remove}also known as Loanable Funds Markets\commb{end remove} 
are a class of DeFi applications which allow users to lend and borrow cryptoassets.
%\commb{removed: on a virtual market} 
At the time of writing, LPs are the most used DeFi applications, 
with the majority of them being deployed on Ethereum \cite{defipulse2021web}. 
% \commb{remove the whole sentence (overlaps with the subsequent ones):} \comal{which one the previous one or the next one?}
% bart: removed this sentence: In LPs, assets are not directly borrowed and lent by agents. 
Deposited funds are pooled and lent on-demand to borrowers, 
only if they possess enough collateral (i.e.\, only if their account is overcollaterized). 
%\commb{I propose to remove the reference to Ethereum: pseudonyms are common to all the public blockchains I am aware of}
As blockchains typically do not provide strong identities, but pseudonyms \cite{buterin2013white}, users' actions are difficult to be regulated under a jurisdiction, which makes collateralization the main protection mechanism against adversarial behaviours \cite{perez2020liquidations}: an agent can only borrow a quantity of tokens worth less than the amount of collateral they deposited. This mechanism and others (e.g.\, interest rates) is in place in order to incentivize borrowers to repay their loans.

We now recall details of the lending pools model in~\cite{bartoletti2020sok}. 
The basic components of the model are \textit{agents} and \textit{cryptoassets}. LP agents are the rational entities taking part in the protocol. Contrarily, LP cryptoassets are token types, each representing a different virtual currency. The model distinguishes two classes of token types: \textit{free tokens} and \textit{minted tokens}, denoted respectively by the sets $\mathcal{T}_{f} = \{\tau_{i}\}_{i \in \lbrack 1..k \rbrack}$
%
%\comtj{For integer intervals, one often sees a version with 2 dots, e.g. $[1..k]$}
%
and $\mathcal{T}_{m} = \{\tau'_{i}\}_{i \in \lbrack 1..k \rbrack}$, where $k$ is the number of cryptocurrencies available on the pool. 
The  difference between these classes of token types is that free tokens have a value established by external markets, whereas minted tokens are assets coined by the protocol, hence holding value only in a specific LP environment. In other words, minted tokens are loyalty credits held by the agents actively joining the protocol. In fact, minted tokens are granted by the protocol to the agents in return for free tokens, hence each minted token $\tau'$ corresponds to a free token $\tau$, also called its underlying token\footnote{The underlying token of $\tau'$ is also denoted as $u_{\pi}(\tau') = \tau$.}.
We denote by $\mathcal{T}$ the set of all token types, i.e.\ $\mathcal{T}\defineas\mathcal{T}_f \cup \mathcal{T}_m$.

\iftrue % TJ: a new proposal
%\comtj{Reshape the following paragraphs, the original ones can be found in an  {\textbackslash}iffalse-block in the source}
Given agents and assets, the LP model yields as a transition system where each state $\Gamma$ is of the form $\Gamma \defineas \sigma\ |\ \pi\ |\ p$:
\begin{enumerate}
\item The \textit{wallets} function $\sigma : \mathcal{A} \rightarrow (\mathcal{T} \rightarrow \mathbb{R}^{+}_{0})$ stores each agent's balance of 
%\comtj{Remove ``free'' as minted tokens are also stored in $\sigma$?}
tokens. For instance, the wallet of an agent $\Agent{A}$ is expressed by the partial function $\sigma_{\Agent{A}}$, and the balance of its $\tau$-typed tokens by $\sigma_{\Agent{A}}(\tau)$.
\item The \textit{pool} component $\pi$ is a triple $(\pi_{f},\pi_{l}, \pi_{m})$. It is composed by three partial functions: $\pi_{f}  : \mathcal{T}_{f} \rightarrow \mathbb{R}^{+}_{0}$ storing the amount of free tokens deposited in the pool, $\pi_{l} : \mathcal{A} \rightarrow (\mathcal{T}_{f} \rightarrow \mathbb{R}^{+}_{0})  $ memorising the loans each agent owes to the pool and $\pi_{m} : \mathcal{T}_{f} \rightarrow (\mathcal{T}_{m} \times \mathbb{R}^{+}_{0}) $ keeping track of the minted tokens (also called the \textit{collateral} or credits) purchased from the pool.
\item The price function $p : dom(\pi_{f}) \rightarrow \mathbb{R}^{+}_{0}$ stores the price of each free token available in the pool.
\end{enumerate}

%\commb{It is not necessary to introduce this heavy notation to present a single rule! Just drop the notation, and improve the comments about the rule}
%
% For simpler notation, we make some additional definitions. First, 
Given a partial map $f$, we denote by $f\{v/x\}$ the point-wise update of $f$ at the point $x$ to the value $v$.
In order to add and remove tokens in the functions defined above, a partial binary operation $\circ : \mathbb{R}^{+}_{0} \times \mathbb{R}^{+}_{0} \rightarrow \mathbb{R}^{+}_{0}$, such as addition, is extended to them. 
Given a partial map $f : \Tokens \rightarrow \mathbb{R}^{+}_{0}$, a token type $\tau \in \Tokens$ and a value $v \in \mathbb{R}^{+}_{0}$, the partial map $f \circ v : \tau$ is defined as
\begin{equation*}
  %\label{eq:circ-def}  
  f \circ v : \tau \defineas
  \begin{cases}
    f\{f(\tau) \circ v/\tau\} \text{ if }\tau \in \mathop{dom}(f) \text{ and } f(\tau) \circ v\text{ is defined}\\
    f\{v/\tau\} \mathrm{\ if\ }\tau \notin \mathrm{dom(}f\mathrm{)}
  \end{cases}
\end{equation*}
\fi
% The original version

In order to describe the model evolution, some additional definitions shall be given. 
The following LP components may rely on the whole state $\Gamma$, or some of its components. 
This dependency is indicated by the means of subscripts. 
For instance, writing $F_{X}$ means that $F$ depends on the $X$ component of the state.

% Firstly, the concept of loan and minted value held by a specific user is explained. 
% Intuitively, the value functions return the current value of respectively the minted tokens owned and the free tokens borrowed by an agent. 
The functions $V_{\Gamma}^{l}$ and $V_{\Gamma}^{m}$ define, respectively,
value of tokens lent to a given agent, and the value of minted tokens owned by a given agent:
% are defined in \Cref{eq:V^m,eq:V^l}\footnote{$\mathit{ER_{\pi}}(\tau, \tau')$ is the exchange rate of the minted token $\tau'$ into $\tau$. 
% For brevity, this is not defined, the interested reader can refer to \cite{bartoletti2020sok}, Section 3.4.
% }.
%\begin{equation}
\[
%  \label{eq:V^l}
  V_{\Gamma}^{l}(\Agent{A}) \defineas \sum_{\tau \in \mathcal{T}_{f}}(\pi_{l} (\Agent{A}))(\tau) \cdot p(\tau)
%\end{equation}
\hspace{0.5cm}
%\begin{equation}
  \label{eq:V^m}
  V_{\Gamma}^{m}(\Agent{A}) \defineas \sum_{\tau' \in \mathcal{T}_{m}} \sigma_{\Agent{A}}(\tau') \cdot \mathit{ER}_{\pi}(\tau', \tau) \cdot p(\tau)
 \]
%\end{equation}

\noindent 
where $\mathit{ER}_{\pi}(\tau', \tau)$ is the exchange rate of minted tokens 
%\comtj{Added reference}
(see Section 3.1 of \cite{bartoletti2020sok}). 

The collateralization of an agent $\Agent{A}$ is defined as  
$C_{\Gamma}(\Agent{A}) = \nicefrac{V_{\Gamma}^{m}(\Agent{A})}{V_{\Gamma}^{l}(\mathrm{\Agent{A}})}$.
This is an essential indicator of agents' lending safety:
in fact, a collateralization below a given threshold ($\CMin$) entails an agent to be liquidated and hence to incur in a financial loss, 
as detailed later. 

The behaviour of agents interacting with a lending pool is formalized as a set of rewriting rules,
which define transitions between states.
Such transitions are written as $\Gamma \xrightarrow{\mathrm{r_{\Agent{A}}}(z^{n})} \Gamma'$,
where $\Gamma$ is the starting state, $\Gamma'$ is the target state, and $\mathrm{r}_{\Agent{A}}(z^{n})$ is the action
(fired by $\Agent{A}$) which triggers the state transition.
Actions have the form $\mathrm{r}_{\Agent{A}}(z^{n})$, where $\mathrm{r}$ is the action name, 
and $z^{n}$ is an $n$-tuple of parameters.

\begin{table}[t]
  \centering
  \begin{tabularx}{\textwidth}{ l X }
    \hline \\ 
    %$\mathrm{Trf}_{\Agent{A}}(\Agent{B}, v: \tau)$ & $\Agent{A}$ transfers  $v$ free-tokens of type $\tau$ from its wallet to $\Agent{B}$'s one.
    %\\ %\RSep
    %$\mathrm{Mtrf}_{\Agent{A}}(\Agent{B}, v: \tau')$ & $\Agent{A}$ transfers $v$ units of minted token $\tau'$ to $\Agent{B}$, as long as $\Agent{A}$ remains overcollateralized.
    %\\ %\RSep
    $\mathrm{Dep}_{\Agent{A}}(v: \tau)$ & $\Agent{A}$ deposits  $v$ free-tokens of type $\tau$ from its wallet to the pool. Subsequently, the pool coins $v'$ units of $\tau'$, with $v'$ computed so to incentivize deposits only if the LP is lacking free tokens.
    \\ %\RSep
    $\mathrm{Rdm}_{\Agent{A}}(v: \tau')$ & $\Agent{A}$ redeems $v$ units of the minted token $\tau'$, as long as $\Agent{A}$'s collateralization is greater than a threshold ($\CMin$) and LP holds enough tokens of type $\tau'$.
    \\ %\RSep
    % As a result $\pi_{f}$ grows by $v$ and $v'$ minted tokens of type $\tau'$ are added to $\pi_{m}(A)$.
    $\mathrm{Bor}_{\Agent{A}}(v: \tau)$ & $\Agent{A}$ borrows $v$ units of a free token $\tau$, assuming it has enough collateral.
    \\ %\RSep
    $\mathrm{Rep}_{\Agent{A}}(v: \tau)$ & $\Agent{A}$ repays $v$ units of its loan in the free token $\tau$ to the LP.
    \\ %\RSep
    $\mathrm{Liq}_{\Agent{A}}(\Agent{B}, v: \hat{\tau}, \tau')$ & $\Agent{A}$ (liquidator) liquidates a variable amount of $\Agent{B}$'s (borrower's) minted tokens $\tau'$, by paying $v$ units of free tokens $\hat{\tau}$. Notably, $\hat{\tau} \in \FTokens$ is in general different from $\tau$, the underlying token of $\tau' \in \MTokens$. This action can be executed only if the $\Agent{B}$'s collateralization is below $\CMin$, meaning $\Agent{B}$ is undercollaterized.
    \\ %\RSep
    $\mathrm{Int}$ & The LP contract accrues interest on the existing loans. %\commb{wasn't the effect of Int that of accruing interests on loans? The current comment suggests that the interest rate is updated, which seems different to me} 
    This disincentivizes borrowers from postponing their loans repayment.
    \\ %\RSep
    $\mathrm{Price}$ & Token prices are updated according to a given price evolution model.
    \VSkip \\ \hline
  \end{tabularx}
  \caption{Summary of some of the lending pools actions from \cite{bartoletti2020sok}.  %\commb{proposal: remove the Trf and Mtrf actions, and add the price update action (note that price updates are used heavily in the paper). Further, what is the point in introducing the whole set of LP actions, if in the paper we only use few of them (perhaps, only Liq)?}
  }
  \label{tab:lp-sum-actions}
\end{table}

The main actions of lending pools are informally summarised in \Cref{tab:lp-sum-actions}. 
Since the focus in this paper is on liquidations as one of the key incentive mechanisms, we will provide the details for such action only.
~\Cref{fig:liquidate-rl} provides a formal description of the rule. 
The essential preconditions to understand the rule are \LiqC{4}, \LiqC{8}, \LiqC{9}, \LiqC{10} and \LiqC{11}.
%
%\comtj{In $\LiqC4$, $\Underlying{\tau'}$ is undefined, change to $\tau$?}
%
%\comtj{Should $\LiqC{7}$ be $\PoolLoansP \defineas \PoolLoans\{\PoolLoansOf{\Agent{B}} - v : \hat{\tau} / \Agent{B}\}$ in the current notation?}

\begin{figure}[h!]
  \[
    \begin{array}{l@{\ }l@{\qquad}l@{\ }l@{\qquad}l@{\ }l}
    \LiqC{1} & \tau' \in \MTokens
    &    
    \LiqC{2} & \WalletOfT{\Agent{A}}{\hat{\tau}} \ge v
    &
    \LiqC{3} & \LoanOf{\Agent{B}}{\hat{\tau}} \cdot \Maxliq \ge v
    \\
    \LiqC{4} & v' \defineas v \cdot \frac{\PriceOf{\hat{\tau}}}{\PriceOf{\tau}} \cdot \Rliq
    &
    \LiqC{5} & \WalletOfT{\Agent{B}}{\tau'} \ge v'
    &
    \LiqC{6} & \PoolFundsP \defineas \PoolFunds + v : \hat{\tau}    \\
    \LiqC{7} & 
    \PoolLoansP \defineas \PoolLoans\{\PoolLoansOf{(\Agent{B})} - v : \hat{\tau} / \Agent{B}\}
    %\PoolLoansOf{\Agent{B}} - v : \hat{\tau}    
    &
    \LiqC{8} & \WalletPOf{\Agent{A}} \defineas \WalletOf{\Agent{A}} - v : \hat{\tau}+ v' : \tau'
    &
      \LiqC{9} & \WalletPOf{\Agent{B}} \defineas \WalletOf{\Agent{B}} - v' : \tau'
    \\
    \LiqC{10} & \CollOf{\Wallet}{\Pool}{\Price}{\Agent{B}} < \CMin
    &
    \LiqC{11} & \CollOf{\WalletP}{\PoolP}{\Price}{\Agent{B}} \le \CMin
 \\    
    \hline 
    \multicolumn{6}{c}{
      \ConfObjOf{\Wallet}{\Pool}{\Price}
      \BTrans{\LiqRule{\Agent{A}}{\Agent{B}}{v:\hat{\tau}}{\tau'}}
      \ConfObjOf{\FSubst{\FSubst{\Wallet}{\WalletPOf{\Agent{A}}}{\Agent{A}}}{\WalletPOf{\Agent{B}}}{\Agent{B}}}
              {\PoolFrom{\PoolFundsP}{\PoolLoansP}{\PoolMint}}
              {\Price}
    }
  \end{array}
  \]
  \caption{The rule for liquidation.}
  \label{fig:liquidate-rl}
\end{figure}
 
 \begin{itemize}
 \item [\LiqC{3}] The amount of repayable loan is limited by a percentage factor $\Maxliq$, as done in 
 Aave \cite{boado2021aave-maxliq} and Compound \cite{peterins2021compound-maxliq}. 
 \item [\LiqC{4}] computes the reward for the liquidating agent. This is based on the liquidated amount $v$ and the reward factor $\Rliq$. The idea is that $\Agent{A}$, by repaying part of $\Agent{B}$'s loan, is reducing the likelihood of the protocol to become illiquid. This behaviour is incentivized by the platform by setting the aforementioned reward to a value strictly higher than 1. A common value for $\Rliq$ is 1.1.
 %\comtj{Remove the `` - '' in {\textbackslash}items so that this becomes ``$\LiqC1$ computes ....'' and so on?}
 \item[\LiqC{8}, \LiqC{9}] update the involved agents' wallets, $\Agent{A}$ repays $v$ units of $\Agent{B}$'s loan in $\hat{\tau}$ and is compensated with $v'$ units of $\tau'$   
 \item[\LiqC{10}] ensures that the rule is executable only if $\Agent{B}$'s collateralization is less than $\CMin$, which is often set to 1.5. This rule is the reason why agents' collateralization should be at least $\CMin$, so to avert the risk of being liquidated and incurring in the loss of the liquidation reward $\Rliq$.
 \item[\LiqC{11}] prevents $\Agent{A}$ from seizing a higher collateral amount than the one required for $\Agent{B}$ to be considered safe (i.e.\ $\Coll(\Agent{B}) \geq \CMin$). 
%\commb{Since the implemented model uses this $\Maxliq$, we should add a premise to the rule}
 \end{itemize}

% \begin{figure}
%   \centering
%   \includegraphics[width=0.9\linewidth]{liquidate-rl.png}
%   \caption{LP liquidate rule. Adapted from \cite{bartoletti2020sok}.}
%   \label{fig:liquidate-rl}
% \end{figure}

\setlength{\intextsep}{0.25cm}
\setlength{\textfloatsep}{0.25cm}
\setlength{\abovecaptionskip}{-0.0cm}

\begin{figure}
  \centering
  \includegraphics[width=0.7\linewidth]{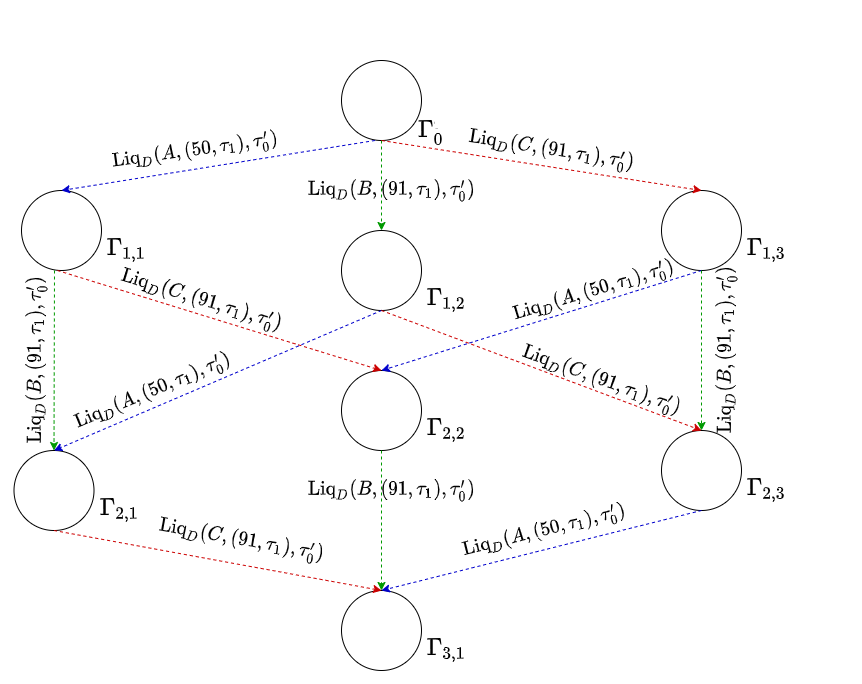}
  \caption{Example transition system.
  %\comav{Text is a bit too small. We should make it bigger (maybe as a normal figure)}
  % produced by the initial state ($\ConfObj_{0}$) holding three liquidate actions, executed by agent $\Agent{D}$. The state of the graph are defined in \Cref{tab:run-ex}. 
  }
  \label{fig:run-ex-ct}
\end{figure}

\Cref{fig:run-ex-ct} illustrates the transition system for a simple running example, where three liquidate actions are executed. 
The figure shows six possible traces all originating from $\Gamma_{0}$ and having $\Gamma_{3,1}$ as final state. 
%\commb{Unclear the role of the index $i$: if not necessary, just call the state $\Gamma_0$}
%
Each state in the figure is defined by a row in \Cref{tab:run-ex}. 
Additionally, transitions, namely $\mathrm{Liq}$ actions performed by $\Agent{D}$, 
are indicated by different colours depending on the liquidated borrower in both the transition system and the table. 
Notably, assuming $\CMin = 1.5$ and $\Rliq = 1.1$, all borrowers in $\ConfObj_{0}$, 
%
%\commb{before you removed the index $i$, so perhaphs you want to remove it also here and below for coherence}
%
$\Agent{A}\mathrm{,}\,\Agent{B}\,\mathrm{and}\,\Agent{C}$, are undercollaterized. 
Specifically, $\Agent{A}$ is marginally undercollaterized since $\CollCustom{\ConfObj_{0}}(A) = 1.25 > 1.1 = \Rliq$, while $\Agent{B}$ and $\Agent{C}$ are strongly undercollaterized, being both $\CollCustom{\ConfObj_{0}}(B)$ and $\CollCustom{\ConfObj_{0}}(C)$ below $1.1$. This allows $\Agent{D}$ to seize the entire $\Agent{B}\ \mathrm{and}\ \Agent{C}'s$ collateral, as evident from $\ConfObj_{3,1}$ in \Cref{tab:run-ex}. Contrarily $\Agent{A}$'s collateralization is restored to $\CMin$.

%\begin{wrapfigure}{r}{0.6\textwidth}
%\end{wrapfigure}

As an example, consider transition $\ConfObj_{0} \xrightarrow{\mathrm{Liq}_{\Agent{D}}(\Agent{B}, 91 : \tau_{1}, \tau_{0}')} \Gamma_{1,2}$. Agent $\Agent{D}$ repays $91$ units of $\tau_{1}$, seizing $91 \cdot \Rliq \approx 100$ units of $\tau_{0}'$ from agent $\Agent{B}$. This also affects $\pi$, in a way that the funds in $\tau_{1}$ are incremented by $91$ units, as illustrated by $\pi_{f}(\tau_{1})$, while $\Agent{B}$'s loan is decremented by $91$ units, as shown by $\pi_{l}(\Agent{B})(\tau_{1})$. Contrarily, $\pi_{m}$ is not modified by the transaction, as the $100$ units of minted tokens $\tau_{0}'$ are simply transferred from $\Agent{B}$'s wallet to $\Agent{D}$'s one.

% Finally, it should be noted that the choice of $\CMin = 1.5$ and $\Rliq = 1.1$ is the default choice in \cite{bartoletti2020sok}. 
% \todo[inline]{Use this example in:\\
%   1. Chap 4: when describing 3 scenarios;\\
%   2. Chap 5: example of search.
% }

% \BeforeBeginEnvironment{array}{\small}
\begin{table}[t]
  \begin{center}
    % \resizebox{\textwidth}{!}{
      $\begin{array}{
        V{\TTick}cV{\TTick}
           cV{\TTick}
           c|c|cV{\TTick}
           % c|cV{\TTick}        
           c|cV{\TTick} 
           c|cV{\TTick}
           c|cV{\TTick}
           c|c|cV{\TTick}
          % cV{\TTick}
          c|c|cV{\TTick}}
          \hline
           \MultiCRFirst{1}{2}{\ConfObj}
           & \MultiCR{1}{2}{\PoolFunds}
           & \multicolumn{3}{cV{\TTick}}{\PoolLoans}
           % & \MultiCR{2}{2}{\PoolMint}          
           % & \MultiCR{2}{2}{\Price}
           & \MultiCR{2}{2}{\WalletOf{\Agent{A}}}
           & \MultiCR{2}{2}{\WalletOf{\Agent{B}}}
           & \MultiCR{2}{2}{\WalletOf{\Agent{C}}}
           & \MultiCR{3}{2}{\WalletOf{\Agent{D}}}
           & \MultiCR{3}{2}{\Coll}
          \\ \cline{3-5}

          & \MultiC{1}{}
           & \Agent{A} & \Agent{B} & \Agent{C}
           % & \MultiC{2}{}                                     
           & \MultiC{2}{}
           & \MultiC{2}{}
           & \MultiC{2}{}
           & \MultiC{3}{}
           % & \MultiC{2}{}          
           & \MultiC{3}{}                                  
           % \\ \cline{2-25}
           \\ \cline{2-17}        

          & \tau_{1}
           & \tau_{1} & \tau_{1} & \tau_{1}
          % & \tau_{0} & \tau_{1}                                                                   
          & \tau_{1} & \tau_{0}' 
          & \tau_{1} & \tau_{0}' 
          & \tau_{1} & \tau_{0}' 
            & \tau_{1} & \tau_{0}' & \tau_{1}'
          % & \tau_{0},\,\tau_{1}
           & \Agent{A} & \Agent{B} & \Agent{C}
          \\ \hline

          %  \ConfObj_0^i
          %  & 195
          %  & 80 & 100 & 125
          %  & 2 & 1                        
          %  & 80 & 100
          %  & 100 & 100
          %  & 125 & 100
          %  & 500 & 0 & 500
          %  & 2.5 & 2 & 1.6
          % \\ \hline
          
           \ConfObj_0^i
          & 195
           & 80 & 100 & 125
           % & 1 & 1                        
          & 80 & 100
          & 100 & 100
          & 125 & 100
           & 500 & 0 & 500
           & 1.25 & 1 & 0.8
          \\ \hline
          
           \ConfObj_{1,1}
          & \Blue{\mathbf{245}}
           & \Blue{\mathbf{30}} & 100 & 125
           % & 1 & 1                                                    
          & 80 & \Blue{\mathbf{45}}
          & 100 & 100
          & 125 & 100
            & \Blue{\mathbf{450}} & \Blue{\mathbf{55}} & 500
          % & 1
            & \Blue{\mathbf{1.5}} & 1 & 0.8
          \\ \hline

           \ConfObj_{1,2}
          & \Green{\mathbf{286}}
           & 80 & \Green{\mathbf{9}} & 125
          % & 1 & 1                                                   
          & 80 & 100
          & 100 & \Green{\mathbf{0}}
          & 125 & 100
            & \Green{\mathbf{410}} & \Green{\mathbf{100}} & 500
          % & 1
           & 1.25 & \Green{\mathbf{0}} & 0.8
          \\ \hline
          
           \ConfObj_{1,3}
          & \Red{\mathbf{286}}
           & 80 & 100 & \Red{\mathbf{34}}
           % & 1 & 1                                    
          & 80 & 100
          & 100 & 100
          & 125 & \Red{\mathbf{0}}
            & \Red{\mathbf{410}} & \Red{\mathbf{100}} & 500
          % & 1
           & 1.25 & 1 & \Red{\mathbf{0}}
           \\ \hline
       \ConfObj_{2,1}
           & \mathbf{336}
           & \Blue{\mathbf{30}} & \Green{\mathbf{9}} & 125
           % & 1 & 1
           & 80 & \Blue{\mathbf{45}}
           & 100 & \Green{\mathbf{0}}
           & 125 & 100
           & \mathbf{359} & \mathbf{155} & 500
           & \Blue{\mathbf{1.5}} & \Green{\mathbf{0}} & 0.8 
        \\ \hline
                
       \ConfObj_{2,2}
        & \mathbf{336}
        & \Blue{\mathbf{30}} & 100 & \Red{\mathbf{34}}
        % & 1 & 1 
        & 80 & \Blue{\mathbf{45}}
        & 100 & 100
        & 125 & \Red{\mathbf{0}}
        & \mathbf{359} & \mathbf{155} & 500
        & \Blue{\mathbf{1.5}} & 1 & \Red{\mathbf{0}}
        \\ \hline

       \ConfObj_{2,3}
        & \mathbf{377}
        & 80 & \Green{\mathbf{9}} & \Red{\mathbf{34}}
        % & 1 & 1 
        & 80 & 100
        & 100 & \Green{\mathbf{0}}
        & 125 & \Red{\mathbf{0}}
        & \mathbf{318} & \mathbf{200} & 500
        & 1.25 & \Green{\mathbf{0}} & \Red{\mathbf{0}}
        \\ \hline
        
       \ConfObj_{3,1}
        & \mathbf{427}
        & \Blue{\mathbf{30}} & \Green{\mathbf{9}} & \Red{\mathbf{34}}
        % & 1 & 1
        & 80 & \Blue{\mathbf{45}}
        & 100 & \Green{\mathbf{0}}
        & 125 & \Red{\mathbf{0}}
        & \mathbf{268} & \mathbf{255} & 500
        & \Blue{\mathbf{1.5}} & \Green{\mathbf{0}} & \Red{\mathbf{0}}
        \\ \hline                    
        \end{array}$
    \caption{States of the transition system in \Cref{fig:run-ex-ct}. For simplicity, the price function $p$ is assumed to be constant such that $p(\tau_{0})=p(\tau_{1}) = 1$ in every state. The values of the LP parameters are $\CMin = 1.5$, $\Rliq = 1.1$ and $\Maxliq = 1$.}
  \label{tab:run-ex}    
  \end{center}
\end{table}

\begin{wrapfigure}{r}{0.5\textwidth}
  %\centering
  %\vspace{-1.0cm}}
  \hspace{-1.0cm}
  \includegraphics[width=0.6\textwidth]{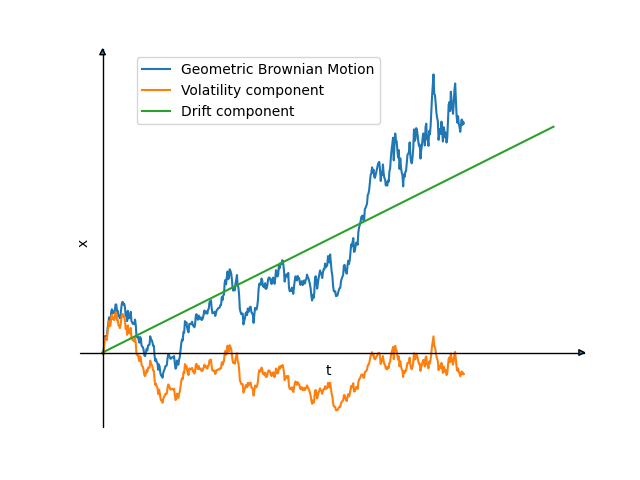}
  \caption{GBM components.
  %\comav{I made it a little bit bigger}
  }
  \label{fig:gbm-comps}
\end{wrapfigure}

\subsubsection{Stock Market Price Modelling}
\label{sec:bg-stock}
We use the \textit{geometric Brownian motion} (GBM) to define a predictive model for price evolution based on past stock market trends. 
A GBM is a continuous-time stochastic process 
%
%\begin{equation}
$
  \label{eq:gbm}
  P_{t} = P_{0} \cdot exp\left[\left(\mu - \frac{\sigma^{2}}{2}\right)t + \sigma W_{t}\right]
%\end{equation}
$.
The two constants $\mu$ and $\sigma$ are respectively called \textit{drift} and \textit{volatility}, whereas $W_{t}$ is a random variable following a \textit{Weiner process}, i.e. a process $ W_{t} \defineas \epsilon \sqrt{dt}$ satisfying the following properties:
(i) $\epsilon \sim N(0,1)$\footnote{It denotes that $\epsilon$ follows a standard normal distribution.} and (ii) for any given pair $(t_{0},t_{0}')$, $W_{t_{0}}$ and $W_{t_{0}'}$ are independent. In other words, a $W_{t}$ is the component yielding the stochastic behaviour of a GBM.
The geometric Brownian motion as a whole can be viewed as the harmonic result of its two components \cite{hull2003options}:
(i) the drift $\left(\mu - \frac{\sigma^{2}}{2}\right)t$ and (ii) the volatility $\sigma W_{t}$. 
The effects of the two  components on the resulting process is shown in \Cref{fig:gbm-comps}. The drift component defines the trend of the resulting process, whereas the volatility component is a measure of the randomly sampled shocks. Intuitively, this signifies that negative values for $\mu$ yield to a downward prediction trend, whereas positive ones to a growth. Oppositely, the higher the $\sigma$ is, the more significantly the prices predictions change.  
Ususall, $\mu$ and $\sigma$ are estimated based on the daily log returns of the targeted stock market~\cite{dmouj2006stock,hull2003options}. Given the closing prices of two consecutive trading days $C_{1}\ \mathrm{and}\ C_{2}$, the log return w.r.t.\,the second trading day is defined as $ln(C_{2}) - ln(C_{1})$.

\section{An LP Simulator for Liquidating Agents}
% \ChapSec{Towards statistical model checking}
\label{chap:agents}

We now lay the foundations for tackling a significant research problem for LPs: finding optimal $\CMin$ and $\Rliq$ parameters. 
%\commb{Changed font compared to the previous section!}
This is achieved by instantiating the LP simulator to conduct statistical analyses of the model. The simulator comprises:
the Maude specification of LPs~\cite{mirelli2021maudelp};
a strategy for automating the behaviour of rational liquidators (\Cref{sec:liq});
and a price evolution model for the three most widely employed cryptocurrencies (\Cref{sec:price}). 
%and the integration of the resulting LP simulator with MultiVeStA (\Cref{sec:multivesta-int}).

\subsection{A Fully-automated Liquidating Strategy}
\label{sec:liq}

This section introduces a liquidating strategy causing the LP protocol to possibly reach unsafe states, where loans are not guaranteed to be repaid. 
We first give an intuitive understanding of aggressive liquidating behaviours, and then describe the proposed liquidating strategy. 
%\commb{stick to a single term: strategy or strategy (or: strategy)}

% Finally, \Cref{subsec:challenges} summarises the main challenges in modelling lending pools and how they have been overcome [see todo there].

% \SSecSSSec{The challenges of modelling LP}
% \label{subsec:challenges}

\subsubsection{The impact of liquidations on collateralization}
\label{subsec:liq-coll}
% 0. overview what are we modelling
% 1. figure -> intuition

Liquidate actions involve two agents: a liquidator, i.e.\ as an agent with enough tokens to fire liquidate actions, 
and a borrower with a collateralization below the threshold $\CMin$.

\begin{wrapfigure}{r}{0.40\textwidth}
  \hspace{-1cm}
  \includegraphics[width=.55\textwidth]{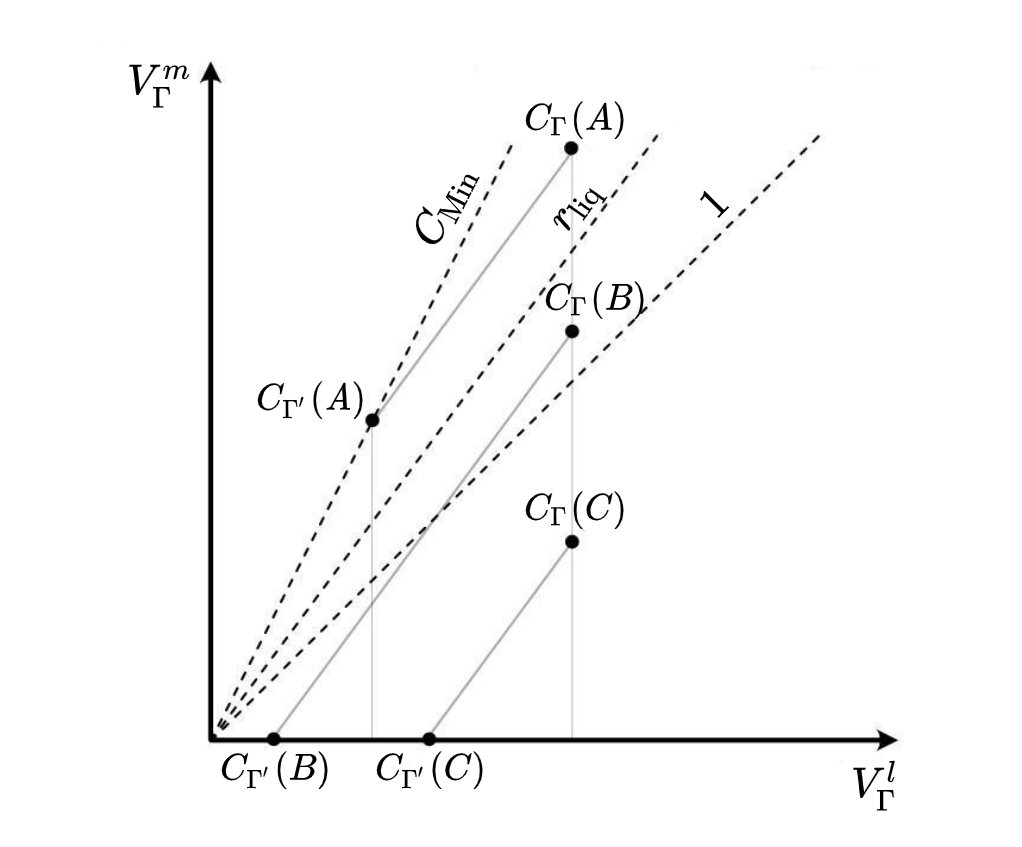}
  \caption{Liquidation scenarios
  %\comav{I made it bigger}
  }
  \label{fig:liq-scenarios}
\end{wrapfigure}

Liquidators have a fundamental role in the financial safety of LPs, 
as they supply free tokens whenever the pool is lacking them. 
% Although they enable an essential mechanism for the LP safety\footnote{Together with interest rates.}, 
On the other hand, excessively zealous liquidators could be harmful to the system,
since they could disincentivize undercollaterized borrowers to repay their loans. 
This is exemplified in~\Cref{fig:liq-scenarios}, where all the liquidating scenarios are outlined. 
The figure illustrates the agents' collateralization\footnote{Defined in \Cref{subsec:lp}.}, detailing the outcomes of liquidate actions in every possible (non-trivial) state. 
The scenarios are also captured by the running example in \Cref{fig:run-ex-ct}.

Firstly, the three dashed lines in the figure correspond to the liquidation parameters specific to the instantiated pool. Their labels represent the respective line slopes. The line labelled $1$ depicts the scenarios where the collateral value equals the loan value\footnote{Given a borrower $\Agent{B}$, collateral value and loan values are defined in \Cref{subsec:lp} as $V^{m}_{\Gamma}(\Agent{B})$ and $V^{l}_{\Gamma}(\Agent{B})$, respectively.}. Consequently, it can be intended as the loan repayment incentivizing threshold, i.e.\,the collateralization value below which borrowing %\comjc{borrowing} 
agents should be considered to be disincentivized in repaying their loans as their outstanding loan debt exceeds their collateral in value. 
%\comjc{as their outstanding loan debt exceeds their collateral in value}.
These residual loans are also called \textit{non-recoverable}.
% \comjc{The loan becomes \textit{non-recoverable} should the collateral value of the borrowing agent reach zero (Agents B/C in state $\Gamma'$ shown in \cref{fig:liq-scenarios}): in such a state, even an increase in the price of the collateral assets cannot incentivize any repayment of the loan.}

Additionally, the three points indicate the initial collateralization of three liquidated borrowers. Each liquidation action\footnote{Or a few of them.} is illustrated by a solid line drawn from $\Coll(\Agent{I})$ to $\CollP(\Agent{I})$ for $\Agent{I} \in \{\Agent{A}, \Agent{B}, \Agent{C}\}$. Liquidations entail a decrease in the liquidated user's collateralization by a linear factor proportional to $\Rliq$  and ultimately determined by the liquidator. Note that the liquidation actions described in the figure follow the semantics of the liquidate action, as the resulting loan value must be greater than zero\footnote{From condition \texttt{(pi).loan[B0][tau] >= v}} and the final collateralization must be at most $\CMin$\footnote{From condition \texttt{(gamma').C(B0) <= CMin}}.

It is worth observing that the liquidations in the figure can be achieved by applying only one action if and only if two conditions hold. Firstly, the liquidator invests enough liquidity to seize the entire seizable collateral. 
Secondly, the liquidated borrower does not diversify the type of the loan. 
If either the first condition or the second does not hold, then the liquidations illustrated in the figure can be achieved uniquely by performing several liquidate actions on the borrower. 
This is frequently the case in the major LP implementations (Compound and Aave). 
In fact, these prevent the whole seizable collateral amount to be atomically liquidated, 
by setting $\Maxliq$ which is variable in Compound \cite{peterins2021compound-maxliq} and constant (equals to 0.5) in Aave \cite{boado2021aave-maxliq}. 
Our model includes the parameter $\Maxliq$ as a constant following Aave, but it could be extended to a variable one. 
%\commb{seems a contradiction: our Maxliq is constant (like in Aave), while we say that in Compound this is variable}

\subsubsection{The proposed liquidating strategy}
\label{subsec:heur}
% 0. motivation
% 1. strategy:
% 1.1. data structures;
% 1.2. estimate of its cost in time
As shown in \Cref{fig:liq-scenarios}, the collateralization of $\Agent{A}$ is re-established, whereas liquidations cause $\Agent{B}$ and $\Agent{C}$ to lose their entire collateral, disincentivizing them from repaying the loans. 
In light of this fact, a relevant research question is whether there exists an optimal pair $(\CMin, \Rliq)$ such that the number of non-recoverable loans is minimal. 

It is worth to observe that, ideally, the closer $\Rliq$ is to 1, the more the collateralization of a loan can drop and still be recoverable by liquidation. Thus a $\Rliq$ marginally greater than 1 is optimal in our model, 
since it would lead to the strongest recovery of user collateralization during liquidation. 
%
%\comjc{...since it would lead to the strongest recovery of user collateralization during liquidation. (there is still a risk of unrecoverable loans, see C in figure 4, but with rliq=1)}
%
However, actual platforms deviate from such ideally optimal $\Rliq$ 
as the costs incurred by liquidators to execute actions have to be compensated by a suitable discount $\Rliq$ on the purchase of minted tokens from the liquidated borrowers.
%
%\commb{...as the costs incurred by liquidators to execute actions have to be compensated by a suitable discount $\Rliq$ on the purchase of minted tokens from the liquidated borrowers} 
%
%\commb{the sentence "Hence [...] Rliq" can be safely removed}
%Hence, it is still interesting to investigate different choices for $\Rliq$. 
%
%In some cases $r_{liq}$ may be fixed so that our research questions consists on finding the optimal $C_{min}$. 
%
In order to investigate the effects of choices  $\Rliq$ and  $\CMin$, we propose a strategy attempting to reproduce a rational behaviour for liquidators. 
The employed strategy simulates a \textit{rational} behaviour where liquidators repay the entire borrowers' loan. 
%\commb{collateral loan
%$\rightarrow$ loan}. 
%\comjc{loan}.
%
The rationality of the behaviour we are going to study is based on the following key observations:

\begin{enumerate}
  
  \item Fast liquidations have the advantage of restoring liquidity whenever the borrowers have collateralization slightly above $\Rliq$  (see agent $\Agent{A}$ in \Cref{fig:liq-scenarios}).
  
  \item On the other hand, fast liquidations may generate non-recoverable loans  whenever the borrowers have collateralization slightly below $\Rliq$ (as for agents $\Agent{B}$ and $\Agent{C}$ in \Cref{fig:liq-scenarios}). 

  \item % \commb{why minimal?}
  Price fluctuations can change the scenario between (1) and (2). 
  For example, it could raise the collateralization of borrowers to $\Rliq$ allowing the liquidators to effectively restore the agents' collateralization to $\CMin$, so that it would be convenient to delay liquidations.
  
\end{enumerate}

%\comjc{The choice of $C_{min}$ and $r_{liq}$ has no effect on the ''speed" of liquidations in our model. The decision to liquidate(+) or not to liquidate(-) is always (+) if the $r_{liq}$ implies a positive liquidation reward. The optimal $r_{liq}$ is marginally larger than 1 and gives the liquidator the minimum possible (positive) reward: any higher liquidator reward lowers the resulting collateralization of the borrower (zero-sum) and is thus sub-optimal. \href{https://docs.google.com/presentation/d/12Q-qdJHprzfdx9ra7L2MpDUCDxypNehgY9J6zaL2MyI/edit?usp=sharing}{See sketch here.} In the real-world, this ``optimal'' $r_{liq}$ may be insufficient, as there are costs to executing actions and rational actors may discount the value of minted/wrapped tokens (we naively value them with the price of the underlying token in our liquidator model).}
%
The strategy used to implement the liquidator behaviour selects the  liquidate input parameters, so to maximise the value of seized collateral. Specifically, given a liquidator $\Agent{L}$, the strategy computes the remaining four parameters of $\mathrm{Liq}$: the borrower's agent identifier ($\Agent{B_{r}}$), the amount of loan to be repaid ($v_{r}$), the type of the asset to be repaid ($\hat{\tau}_{r}$) and the one of the asset to be seized ($\tau_{r}'$). 
Because of space constraints, we refer to~\cite{Mirelli2021} for a detailed account of the strategy. 

%\comjc{My concern here is that we claim to explore optimal $C_{min},r_{liq}$ assuming $r_{liq}$ values in the regime $r_{liq}>1$ have differing effects on the liquidator's decision to liquidate or not to (points 1/2 above). But our liquidation strategy does not exhibit this behaviour. Rather, we only explore the borrower collateralization resulting from choices of $C_{min},r_{liq}$, price behaviour and liquidators which are insensitive to increasing $r_{liq}$ when $r_{liq}>1$. May I suggest that we say that these results help inform the loan platform parameterization once we know which minimal $r_{liq}$ is necessary to incentivize liquidation actions in the real-world (when there are costs to executing actions which must be offset by the reward): here the RQ becomes: given $r_{liq}$, what is optimal $C_{min}$, such that ...}

\subsection{Price Modelling}
\label{sec:price}

This section describes the price model employed to predict cryptocurrencies prices, based on historical data. We start with an overview of the price model to motivate its adoption. Afterwards, we present the three model instantiation scenarios used in the subsequent statistical analysis.

\subsubsection{Predicting cryptocurrency prices}
\label{sec:brown-motion}
% 0. why we need it
% 1. describe the analysis type (based on historical data)
%    and mention it was inspired by gudgeon2020decentralized
% 3. explain the setup (2 assets: price(loan) -> increasing ;
%                                 price(collateral) -> dropping)
%    what currencies were considered and how implemented the
%    negative correlation \cite{hull}

The cryptoassets prices are derived from a statistical model representative of the past price behaviour based on the GBM. A GBM is instantiated by two parameters drift and volatility which can be estimated from the currency historical data. This makes the GBM the ideal stochastic process for modelling stock prices based on their past evolution \cite{dmouj2006stock}.

Aiming at stress-testing the LP protocol and inspired by \cite{gudgeon2020decentralized}, we have designed three different scenarios, each comprising a pair of price trends. In practice, each scenario simulates the evolution of prices of a given collateral and loan assets, in a way that respectively when the former declines, the latter increases. In fact, assuming that each borrower $\Agent{B_{0}}$ owes a loan in only one asset type $\tau_{l}$\footnote{Also called a loan asset.} and similarly holds collateral of only one asset type $\tau_{m}$\footnote{Also called a collateral asset.}, such a model for prices necessarily causes some borrowers to become undercollaterized, as shown in \Cref{eq:C-dropping}.

\begin{equation}
  \label{eq:C-dropping}
  C_{\Gamma}(\Agent{B_{0}}) = \frac{V_{\Gamma}^{m}(\Agent{B_{0}})}{V_{\Gamma}^{l}(\Agent{B_{0}})} \xrightarrow{p(\tau_{m}) \rightarrow 0 \ p(\tau_{l}) \rightarrow V} 0 \mathrm{, with }\  V \gg 0
\end{equation}

% Consequently, such cases could also entail the generation of unrecoverable wallets, in case the agent is initially undercollaterized.
More precisely, prices modelling is achieved by opportunely gathering the data used to estimate the parameters (drift and volatility) for generating a growing, decreasing or relatively constant GBM process. In the literature, daily closing prices of stock markets are used since their samples generally tend to be normal, which allows to employ the GBM generic formula. Ultimately, since prices' predictions pairs should variate in a way that they simultaneously display an opposite behaviour, it is necessary to correlate them, as shown in \cite{hull2003options}.

\subsubsection{Prices model instantiation}
\label{subsec:prices-preds-inst}
% 0. prices datasets from CoinGecko APIs, currencies picked based on:
%    0.0 the daily closed prices;
%    0.1 normality tests.
% 1. Show a plot of a sample dataset (growing vs dropping assets)

Given a collateral asset $\tau_{m}$ and a loan asset $\tau_{l}$, the three prices evolution pairs are shown in \Cref{tab:prices-pairs}.

\begin{table}[h]
  \centering
  \begin{tabular}{|c|c|c|c|c|}
    \hline
    Scenario & $\tau_{m}$ & $\tau_{l}$ & $p(\tau_{m})$ & $p(\tau_{l})$ \\
    \hline    
    ETH-WBTC & ETH & WBTC & Declining & Increasing \\
    \hline    
    ETH-USDC & ETH & USDC & Declining & Constant \\
    \hline    
    USDC-WBTC & USDC & WBTC & Constant & Increasing \\
    \hline
  \end{tabular}
  \caption{The three implemented prices evolution scenarios}
  \label{tab:prices-pairs}
\end{table}

The choice of the cryptocurrencies in the table\footnote{Ethereum (ETH), USD Coins (USDC) and Wrapped Bitcoins (WBTC).} is motivated by their closing price historical evolution in three different trimesters (shown in the Appendix, \Cref{fig:hist-cp}). By using those samples, it is possible to simulate the desired trends indicated in the columns named $p(\tau_{m})$ and $p(\tau_{l})$. This is achieved by estimating the expected price returns ($\mu$) and the price volatility ($\sigma$), which are utilised as the drift and volatility instantiating the resulting GBM. The two parameters are estimated according to \cite{hull2003options}. The drift $\mu$ is simply obtained by computing the mean over the closing prices\footnote{Section 15.3 of \cite{hull2003options}.}. Contrarily, $\sigma$  is calculated as $\sigma = \frac{s}{\sqrt{T}}$,  where $T = \frac{91}{365}$, $s$ indicates the standard deviation of the log returns and $\sqrt{T}$ is the annualisation constant.

%\begin{equation}
%  \label{eq:sigma-est}
%  \sigma = \frac{s}{\sqrt{T}},\ \mathrm{with} \ T = %\frac{91}{365}
%\end{equation}

The selected sampling time span (91 days, i.e.\,a trimester) is motivated by the fact that cryptoassets are subject to sudden fluctuations and, even though short samples might not be representative of the entire population, this is a consolidated practice \cite{hull2003options}. 
Besides, the resulting price predictions span over the same time frames, as each price model instantiation produces 91 prices predictions, as illustrated in \Cref{subsec:prices-mv}. 
Notably, the selected cryptocurrencies (ETH, USDC and WBTC) were among the four-most-utilised assets on the Compound market \cite{compound2021markets} at the moment of writing. 
Lastly, the selected closing price samples are suitable, since the derived log returns distributions tend to be normal.

\begin{wrapfigure}{c}{0.5\textwidth}
%\begin{table}[h]
  \centering
  \begin{tabular}{|c|c|c|c|}
    \hline
    Currency & $\mu$ & $\sigma$ & $P_{0}$ (usd) \\
    \hline    
    ETH & -0.012 & 0.12 & 3269.08 \\
    \hline    
    USDC & -7.84E-5 & 0.005 & 0.99 \\
    \hline    
    WBTC & 0.012 & 0.094 & 57260.0 \\
    \hline
  \end{tabular}
  \caption{GBM parameters}
  \label{tab:gbm-params}
%\end{table}
\end{wrapfigure}

\Cref{tab:gbm-params} shows an estimation of the GBM parameters ( obtained from the close prices in the Appendix, \Cref{fig:hist-cp}), by the previously discussed methodology. The parameters are then utilised to instantiate the six GBM processes (each for price evolution), simulating the scenarios in \Cref{tab:prices-pairs}. Finally, the asset initial price $P_{0}$ is a constant set to the actual price in USD of each asset on May 5th, 2021\footnote{According to \href{https://www.coingecko.com/api/documentations/v3}{CoinGecko APIs}.}.

\subsubsection{Expected price predictions}
\label{subsec:prices-mv}
% 0. show the mquatex property
% 1. shows the resulting graphs
% 2. note that they are negatively correlated according to hull

We have used the MultiVeStA statistical analyzer to examine the prices predictions generated by the GBM in each of the scenarios explained in \Cref{subsec:prices-preds-inst}. 
%
%\begin{mdframed}[style=mq-style,
%                 innerbottommargin=-3.0\baselineskip ]
%  \begin{lstlisting}[
%    mathescape=true,
%    label=lst:meanColl-mq,
%    caption={\texttt{meanCollateralPrice.multiquatex} - MultiQuaTEx property producing the mean collateral asset price evolution.}, 
%    style=mq-file-style
%    ]
%$\mathit{obsAtStep}$(i,x) =
%if ( s.rval($\mathtt{"steps"}$) == x )
%  then s.rval(i)
%  else # $\mathit{obsAtStep}$(i,x)
% fi ;
%eval parametric(E[ $\mathit{obsAtStep}$($\mathtt{"CUR\_COLLATERAL\_PRICE"}$,x) ], 
%              x,1,1,91) ;  
%  \end{lstlisting}
%\end{mdframed}
%
%The main idea of this property is that at each step of a simulation, MultiVeStA questions the current \texttt{LPState} about the new value of the collateral asset price, sampled by a geometric Brownian motion. Subsequently, for each liquidation round the mean of the step-wise observations is computed and returned. This is the semantics expressed by the \texttt{E[}\textit{obsAtStep(...)}\texttt{]} input argument to \texttt{eval parametric}. The remaining arguments to \texttt{eval parametric} repeat the same procedure (\texttt{E[}\textit{obsAtStep(...)}\texttt{]}) for 91 liquidation rounds\footnote{The number 91 is motivated by the historical time spans considered to instantiate the geometric Brownian motion processes, discussed in \Cref{subsec:prices-preds-inst}.}. A similar property has also been developed for querying the mean loan asset price evolution, for 91 liquidation rounds.
%
The details are provided in the appendix (
\Cref{fig:norm-all}), and show the normalised trend of the price scenarios, discussed in \Cref{subsec:prices-preds-inst}. The figures in the appendix show that the expected behaviour, expressed in \Cref{tab:prices-pairs} is obtained in all the considered scenarios. Additionally, in \Cref{fig:norm-eth-wbtc,fig:norm-usdc-wbtc} prices predictions are strongly correlated as it is expected. In fact, the GBMs pairs were instantiated as negatively correlated processes\footnote{With $\rho = 1$, where $\rho$ is the Pearson correlation coefficient.} accordingly to \cite{hull2003options}, Section 14.5. Contrarily, \Cref{fig:norm-eth-usdc} shows less correlated prices predictions. This is probably due to the fact that the computation was bounded to execute maximum $5,010$ simulations. In fact, from experimental evidence, the approximation seem to converge at a very slow speed\footnote{Each observation would take around 2 hours to reach the respective CI.}. 

%Finally, \Cref{fig:unnorm-coll-all,fig:unnorm-loan-all} illustrate the unnormalised behaviour of the considered GBM prediction scenarios, w.r.t.\,the collateral and loan assets respectively.       
\section{Statistical Analysis of Liquidation Scenarios}
\label{chap:tool}

We have experimented with the LP model simulator described in \Cref{chap:agents} in order to answer the question: \emph{given a specific scenario, what is the impact of the pair of LP parameters $\CMin$ and $\Rliq$?}

We have considered scenarios generated by four factors. First, the liquidator logic defined in \Cref{sec:liq}, determines immediate and quick\footnote{The repayed debt amounts to \texttt{Maxliq} $\cdot\ t$, with $t$ total amount of repayable debt.} liquidations, causing a significant financial loss to the liquidated party. Secondly, the agent to be liquidated is selected so to maximise the value of seized collateral, which is the most beneficial and rational option for liquidators. Thirdly, liquidators are assumed to hold an \textit{unbounded} amount of resources, which allows them to repeat liquidations as long as there exists an undercollaterized agent. Finally, cryptoasset prices evolve following a trend aimed at causing borrowers to suddenly become undercollaterized.
%\commb{After this paragraph it is still unclear to me what is an undesirable scenario (and why)}

\begin{wrapfigure}{r}{0.5\textwidth}
  %\centering
  \vspace{-0.5cm}
  \hspace{-1.1cm}
  \includegraphics[width=.65\textwidth]{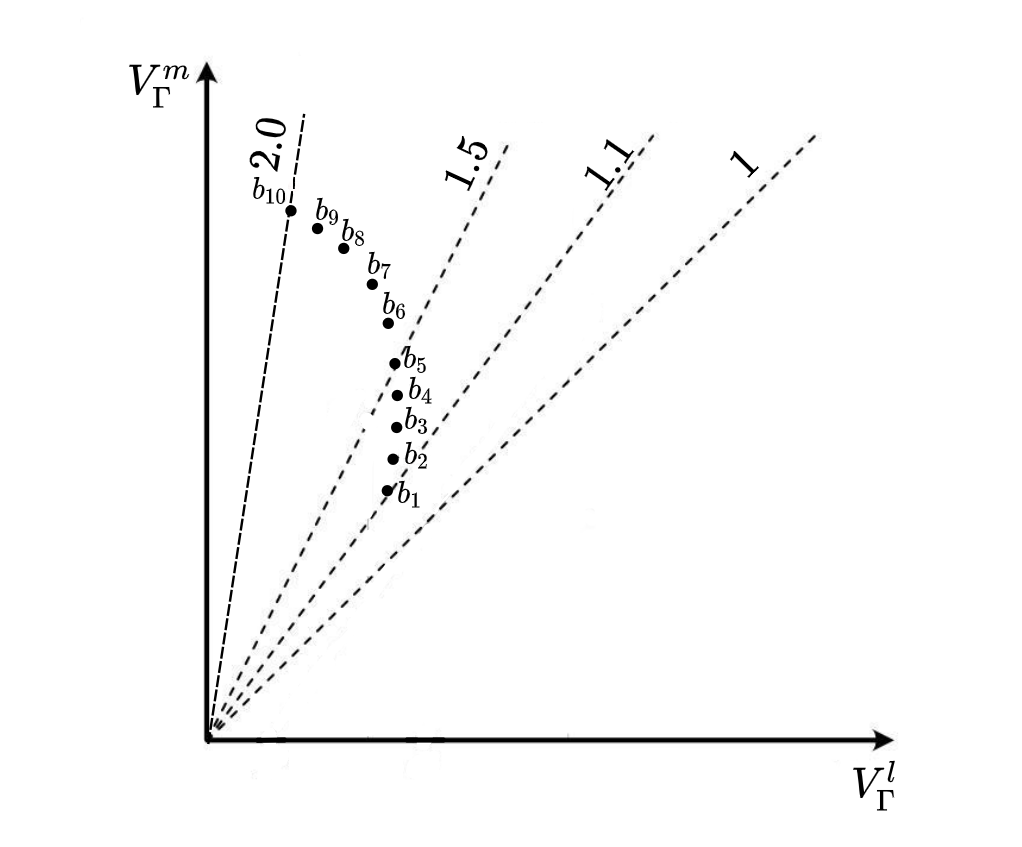}
  \caption{Distribution of collateralization in initial configurations.
  %\comav{I made it bigger}
  }
  \label{fig:c-ag-distr}
\end{wrapfigure}

We recall that the effect of the pair $\CMin$ and $\Rliq$ we are looking for is one that minimises the number of undercollaterized borrowers. 
  %\comjc{I refer to comment and suggestion in liquidator heuristic section. Optimal $r_{liq}$ in our model is marginally greater than 1. Exploration of higher values of $r_{liq}$ assume they have an affect the liquidator's decision to liquidate or not to in regime $r_{liq}>1$, but not in our model.} 
  We have explored the space of choices for the pair by executing MultiVeStA experiments for all $\CMin$ ranging, with step $0.1$, from $1.2$ to $1.5$ and $\Rliq$ ranging from $1.1$ to $\CMin - 0.1$. These ranges were selected based on the values typically assigned to these parameters in the actual implementations: $\CMin = 1.5$ and $\Rliq = 1.1$ \cite{bartoletti2020sok}.

On these premises, we first illustrate the LP model initial configurations used for the subsequent experimentation. Next, we present the results of the performed experiments.

\subsubsection{Initial configurations}
\label{subsec:exp-init-config}

The initial configurations were designed to test the resistance of different borrowers' collateralization to becoming unrecoverable, when subject to repeated liquidations. Since the intention is to observe the model behaviour under three price models (\Cref{subsec:prices-preds-inst}), three different initial configurations are produced, each having a different price for collateral and loan assets. All the configurations share the same amount and types of agents. Specifically, a generic initial configuration comprises ten borrowers having collateralization ranging from 1.0 to 2.0, with step 0.1. This is depicted in \Cref{fig:c-ag-distr}, where $b_{i}$ represents the generic borrower $\Agent{B_{i}}$'s collateralization ($\CollI(\Agent{B_{i}})$), for $\ConfObj^i$ initial configuration. 
Additionally, an arbitrary number of liquidators (three) are added to each configuration. %This determines a race condition at each simulation step, as shown in \Cref{fig:mv-lp-full-ts}.

\subsubsection{Experimental results}
\label{subsec:exp-rst}
%
%\begin{mdframed}[style=mq-style,
%                 innerbottommargin=-3.7\baselineskip ]
%  \begin{lstlisting}[
%    mathescape=true,
%    caption={\texttt{perAgentCollateralization.multiquatex} - MultiQuaTEx property producing the mean collateralization per agent (maximum 10 borrowers).}, 
%    label=lst:meanPerAgColl-mq,
%    style=mq-file-style
%    ]
%$\mathit{obsAtStep}$(i,x) =
%if ( s.rval($\mathtt{"steps"}$) == x )
%  then s.rval(i)
%  else # $\mathit{obsAtStep}$(i,x)
%fi ;
%eval parametric(E[ $\mathit{obsAtStep}$($\mathtt{"1\_COLLATERALIZATION"}$,x) ], 
%                E[ $\mathit{obsAtStep}$($\mathtt{"2\_COLLATERALIZATION"}$,x) ],
%                E[ $\mathit{obsAtStep}$($\mathtt{"3\_COLLATERALIZATION"}$,x) ], 
%                E[ $\mathit{obsAtStep}$($\mathtt{"4\_COLLATERALIZATION"}$,x) ], 
%                E[ $\mathit{obsAtStep}$($\mathtt{"5\_COLLATERALIZATION"}$,x) ], 
%                E[ $\mathit{obsAtStep}$($\mathtt{"6\_COLLATERALIZATION"}$,x) ],  
%                E[ $\mathit{obsAtStep}$($\mathtt{"7\_COLLATERALIZATION"}$,x) ],  
%                E[ $\mathit{obsAtStep}$($\mathtt{"8\_COLLATERALIZATION"}$,x) ],  
%                E[ $\mathit{obsAtStep}$($\mathtt{"9\_COLLATERALIZATION"}$,x) ],  
%                E[ $\mathit{obsAtStep}$($\mathtt{"10\_COLLATERALIZATION"}$,x) ],
%                x,1,1,91) ;
%  \end{lstlisting}
%\end{mdframed}

The results discussed here were obtained by performing MultiVeStA experiments of the LP simulator. 
%The evaluation was performed on an 8GB-RAM machine with two 1.20GHz cores. 
%
Specifically, the inputs to the tool are: the LP simulator discussed in \Cref{chap:agents}, a  MultiQuaTEx property to express the desired measure to be estimated (the expected collateralization value at each liquidation round for each borrower), and a pair of statistical parameters defining the confidence interval (CI) of interest: the maximum confidence interval width $\delta$ and the confidence level $\alpha = 0.05$ which provides 95\% statistical confidence that the estimated value is in the confidence interval. For each property, MultiVeStA will generate enough simulation to meet the required CI. % statistically-valid estimations of the desired measure until the required confidence interval (CI) is reached. 
%The next two paragraphs describe the inputs to MultiVeStA\comav{no info is given on MV inputs - were they removed without updating this sentence?}, afterwards the results are introduced and discussed.

\begin{figure}
%\begin{wrapfigure}{r}{0.5\textwidth}
  \centering
    %\adjincludegraphics[width=0.5\textwidth,trim={0 {0.08\height} 0 {0.15\height}},clip]{l-eth-wbtc-none-ag_1_10.png}
    \adjincludegraphics[width=0.9\textwidth,trim={0 {0.03\height} 0 {0.1\height}},clip]{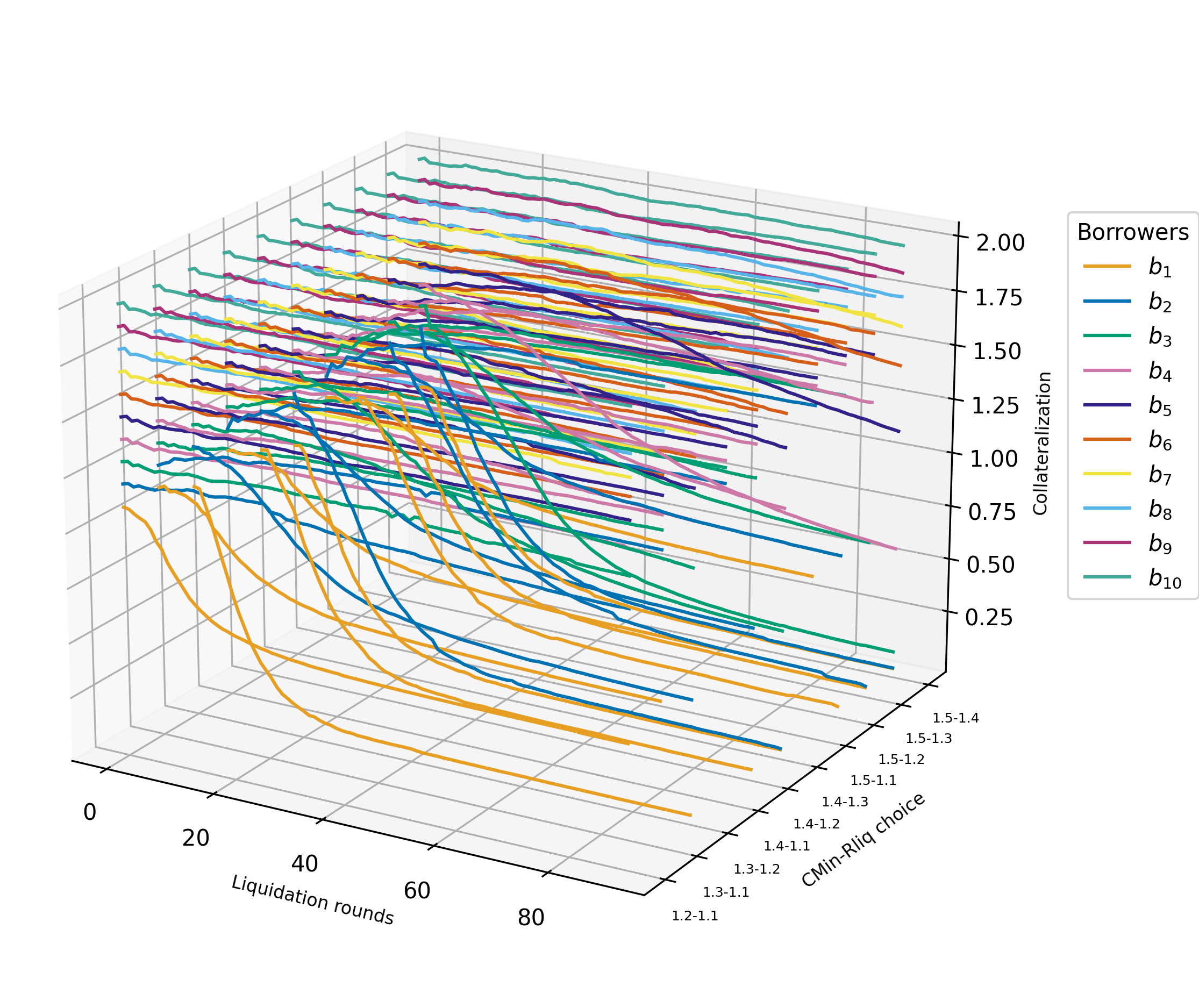}
  \caption{Per-borrower collateralization ($\mathrm{b_{1}\ to\ b_{10}}$) in the ETH-WBTC prices scenario, for varying liquidation rounds and \texttt{CMin-Rliq} choices.
  %\comav{It was ay too small. I changed to figure environment. Fonts should be bigger}
  }
      \label{fig:l-eth-wbtc-none-ag_1_10}
  \label{fig:3d-l2s}
%\end{wrapfigure}
\end{figure}

\Cref{fig:l-eth-wbtc-none-ag_1_10} shows the per-borrower collateralization for varying liquidation rounds and choices of $\CMin$ and $\Rliq$ in the eth-wbtc prices scenario, with a fixed $\Rliq = 1.1$. In this scenario, one can see that undercollaterized agents have a very different behaviour than overcollaterized ones. Specifically, the undercollaterized agents undergo very serious liquidations, which often lead them to unrecoverability, as their collateralization converges to a constant below $\CMin$. Contrarily, overcollaterized agents do not incur in severe financial losses.

Additionally, our experiments (presented in detail in the Appendix, \Cref{fig:s-eth-wbtc-cmin-rliq-ag_1_5,fig:s-eth-usdc-cmin-rliq-ag_1_5,fig:s-usdc-wbtc-cmin-rliq-ag_1_5}) show that the $\CMin$ and $\Rliq$ having the least negative effects on undercollaterized balances is  $\CMin = 1.5$, $\Rliq = 1.1$. This is also quantitatively confirmed by the numbers in \Cref{tab:rst-b-1}. Intuitively, this is a consequence of the fact that when $\CMin=1.5$ and $\Rliq=1.1$ the collateralization of each agent $b_{1}$ to $b_{5}$ is higher on average than for any other $\CMin$ and $\Rliq$ pairs. As a result, the number of unrecoverable loans, the ones held by agents whose collateralization is below $1$, is minimised.

\begin{wrapfigure}{r}{0.55\textwidth}
  \centering
  \begin{tabular}{| l | l | l | l |}
    \hline
    Price scenario & \multicolumn{3}{c|}{\texttt{(CMin-Rliq)}} \\
    \cline{2-4}
                   & (1.5-1.1) & (1.4-1.1) & (1.3-1.1) \\
    \hline
    ETH-WBTC & 0.7115 & 0.6518 & 0.6137 \\
    % \hline
    ETH-USDC & 0.7106 & 0.6583 & 0.6231 \\
    % \hline
    USDC-WBTC & 0.8381 & 0.7739 & 0.7299 \\
    \hline
  \end{tabular}
  \caption{Minimum average $\Coll(\Agent{B_{1}})$}  
  \label{tab:rst-b-1}
\end{wrapfigure}

Finally, our experiments (presented in detail in the Appendix, \Cref{fig:s-eth-wbtc-cmin-rliq-ag_3_7,fig:s-eth-usdc-cmin-rliq-ag_3_7,fig:s-usdc-wbtc-cmin-rliq-ag_3_7}) show that overcollaterized borrowers could still incur in liquidations, in case the prices abruptly change as in the prices scenario ETH-WBTC\footnote{This is mainly determined by the opposite drift and high volatility used to model the ETH and WBTC prices evolution (\Cref{tab:gbm-params}).}. Differently, in the other scenarios, employing the stable coin USDC, overcollaterized agents are, on average, rarely liquidated.

% \Cref{fig:3d-results} shows the collateralization per user, for varying \texttt{CMin-rliq} choices. \Cref{fig:eth-wbtc-cmin-rliq-all,fig:eth-usdc-cmin-rliq-all,fig:usdc-wbtc-cmin-rliq-all} show that the \texttt{CMin-rliq} choice maximising every borrowers' collateralization is $\mathtt{CMin} = 1.5$ and $\mathtt{rliq} = 1.1$. This is also confirmed by \Cref{tab:rst-b-1}, showing the minimum average \texttt{b1}'s collateralization, per liquidation round, for the three \texttt{(CMin-rliq)} choices maximising its collateralization.

% \begin{figure}
%   \centering
%   \begin{subfigure}{0.5\textwidth}
%     \centering
%     \includegraphics[width=\textwidth]{eth-wbtc-cmin-rliq-all.png}
%     \caption{Prices scenario eth-wbtc.}
%     \label{fig:eth-wbtc-cmin-rliq-all}
%   \end{subfigure}
%   \medskip       
%   \begin{subfigure}{0.5\textwidth}
%     \centering
%     \includegraphics[width=\textwidth]{eth-usdc-cmin-rliq-all.png}
%     \caption{Prices scenario eth-usdc.}
%     \label{fig:eth-usdc-cmin-rliq-all}
%   \end{subfigure}
%   \medskip
%   % \end{figure}
%   % \begin{figure}[h] \ContinuedFloat
%   %   \centering
%   \begin{subfigure}{0.5\textwidth}
%     \centering
%     \includegraphics[width=\textwidth]{usdc-wbtc-cmin-rliq-all.png}
%     \caption{Prices scenario usdc-wbtc.}
%     \label{fig:usdc-wbtc-cmin-rliq-all}
%   \end{subfigure}
%   \caption{Average .}
%   \label{fig:3d-results}
% \end{figure}

\section{Related Works}

\label{sec:related-work}
% 0. Imperial college (gudgeon-perez-werner) -> ABS
% 1. Gauntlet -> ABM (make sure to underline this point)
% 2. Tolmach, Bigi...

Verification of DeFi applications is a fairly recent research area where several techniques have been applied. 
%~\cite{liu2019survey,tolmach2021survey}. 
%\commb{It seems to me that the first of these two surveys (liu2019survey) does not cover DeFi at all}
We focus our discussion on works devoted to formal modelling and reasoning of DeFi applications, which typically follow two parallel directions:
%\begin{enumerate}
%\item 
verification of the model properties \cite{bigi2015validation,bai2018formal,abdellatif2018formal,tolmach2021formal}, 
%\item
and statistical analysis of the model variables \cite{angeris2019analysis,chitra2019agent,kaoanalysis,chitra2020stake,gudgeon2020decentralized}. 
%\end{enumerate}

%\comjc{Formal verification SC survey: \cite{tolmach2021survey}}.
%\commb{I have weakened the claim in the next sentence, since there are some papers by Andrychowicz et al. on the formal verification of Bitcoin contracts (also using Uppaal) which date back to 2014.}
The work in~\cite{bigi2015validation} is one of the first addressing formal verification of smart contract properties. 
Their study combines a game-theoretic approach with probabilistic model checking, ultimately validating their results with the model checker PRISM \cite{kwiatkowska2011prism}. 
%\commb{please remove the following citataion (bai2018formal): it is a 5-pages paper which uses SPIN to model check an e-commerce example that they try to sell for a "smart contract". There are more serious works on smart contracts verification around.}
%Using a similar methodology,  \cite{bai2018formal} investigated properties of these systems using the model checker SPIN \cite{holzmann1997model} in order to develop secure templates for smart contracts. 
%
Another example of research in this direction is Tolmach et al.\,\cite{tolmach2021formal} which developed the first multi-pools model and verified invariant properties initially formulated by \cite{bernardi2020wip}. 
Finally, \cite{abdellatif2018formal} proposed a very relevant study on smart contracts, by modelling  not only the contracts and the agents' behaviour but also the underlying blockchain using the BIP framework \cite{basu2011rigorous} and statistical model checking (as we do).  
%
%This is also related to our work as it made use of \textit{Statistical Model Checking} (SMC) \cite{holzmann2018explicit}, a formal verification technique which has strongly influenced the development of MultiVeStA \cite{sebastio2013multivesta,vandin2021automated}.
% In fact, SMC verifies probabilistic properties of a model by using statistical means, i.e.\,Monte-Carlo simulations, whose number is estimated on-the-fly. The estimation is based on two input parameters the risk level, $\alpha$, and the precision, $\delta$. Then, if $p \mathrm{\ and\ } p'$ are, respectively, the real and the estimated probabilities for the property $\phi$ to hold in a given state, simulations are performed until $\mathbb{P}(| p - p' | \leq \delta) \geq 1 - \alpha$ holds \cite{abdellatif2018formal}. Thus, this methodology clearly resembles the technique employed by MultiVeStA, shown in \Cref{subsec:multivesta-intro}. 
%
%The more recent direction of studies on smart contracts \commb{which one?} is also loosely relevant to
The work in 
~\cite{abdellatif2018formal} employs statistical methods too. However, in their case, statistics is useful to estimate unknown variables of the analysed model, hence deriving desirable properties. The quantitative variables estimation is also achieved by performing Monte-Carlo simulations, with a more closely look at  the behaviours displayed by agents \cite{chitra2019agent}. 
In fact, most of this research in this line \cite{kaoanalysis,chitra2020stake,angeris2019analysis} bases its results on Agent-Based Simulations, which is employed to stress test the actual smart contracts implementations being executed on a \emph{``custom-built Ethereum virtual machine that is written in C++''} \cite{kaoanalysis}. 
This research direction, although suggesting promising results, is not ultimately supported by strong statistical guarantees, since the number of Monte Carlo simulations performed to run their analyses is arbitrarily chosen and not backed by a formal justification \cite{kaoanalysis,gudgeon2020decentralized}. 
Nonetheless, a work relevant to ours is the analysis conducted in~\cite{kaoanalysis} on the Compound protocol scalability in face of high stock market prices volatility. 
Similarly to our work, their analysis models the prices by the use of the GBM. 
However, their data collection and analysis methodologies are very different. In fact, they do not sample entire historical periods as illustrated in \Cref{sec:price} for estimating prices volatility. Contrarily, they simply evaluate the minimum and maximum volatility values ever observed and instantiate the GBM for different prices volatilities so to simulate several market environments. 
Finally, the prices model in \Cref{sec:price} has been mostly inspired by \cite{gudgeon2020decentralized}. Similarly to \cite{kaoanalysis}, they stress-test an LP model, not a specific implementation, by using the same price model explained in \Cref{sec:price}. Nonetheless, a remarkable difference is that they instantiate the predictions of the collateral and loan assets pairs with three different correlation parameters. We assume instead predictions of prices pairs to be strongly negatively correlated ($\rho = -1$), in order to simulate the worst-case scenario. Additionally, we reproduced \cite{gudgeon2020decentralized}'s environment by using historical data of three different real cryptoassets on the market.

% In addition, the flexibility of the Maude language and the manifold applications of its verification environment motivate the language choice. 

\section{Conclusions}

We have presented a tool for the analysis of lending pools, an archetypal DeFi application. 
Overall the tool consists of  
%\begin{enumerate}
%\item
(i) an accurate LP simulator based on the model of~\cite{bartoletti2020sok} which can support both the study of vulnerabilities and attacks of LPs;
%\item
(ii) a model checker capable of doing simple reachability analysis and verifying whether LTL properties hold of specific configurations;
%\item
and (iii) a tool for statistical analysis backed by the statistical model checker MultiVeStA. 
%\end{enumerate}
In this paper, we have focused on (iii) and we have shown how to use it to optimize the LP parameters under specific scenarios. 
Details on (i) and (ii) as well as further examples, including reproduction of price oracle attacks using reachability analysis and LTL model checking 
are available in~\cite{Mirelli2021}. 

Future research supported by the developed tool could include the formalization of further attacks and properties of the model. 
Specifically, one could study resistance to illiquidity, as suggested by \cite{kaoanalysis}, or the behaviour of multi-pools configurations, each offering different market opportunities\footnote{In terms of prices and interest rates.} to agents, as proposed by \cite{werner2021sok} and partially developed in \cite{tolmach2021formal}.

\paragraph*{Acknowledgements}
Massimo Bartoletti is partially supported by 
Conv.\ Fondazione di Sardegna \& Atenei Sardi project
F75F21001220007 \emph{ASTRID}. 
James Hsin-yu Chiang is supported by the PhD School of DTU Compute. 
Alberto {Lluch Lafuente} is partially supported by the EU H2020-SU-ICT-03-2018 Project No. 830929 CyberSec4Europe (\href{https://www.cybersec4europe.eu}{cybersec4europe.eu}).
Andrea Vandin is partially supported by the %Independent Research Fund Denmark 
DFF project REDUCTO 9040-00224B.

\bibliographystyle{splncs04}
\bibliography{references/bibliography.bib}   

\clearpage

\appendix

\section{Figures}

\begin{figure}
  \centering
  %\begin{subfigure}{0.32\textwidth}
  \begin{subfigure}{0.6\textwidth}
    \centering
    \includegraphics[width=\textwidth]{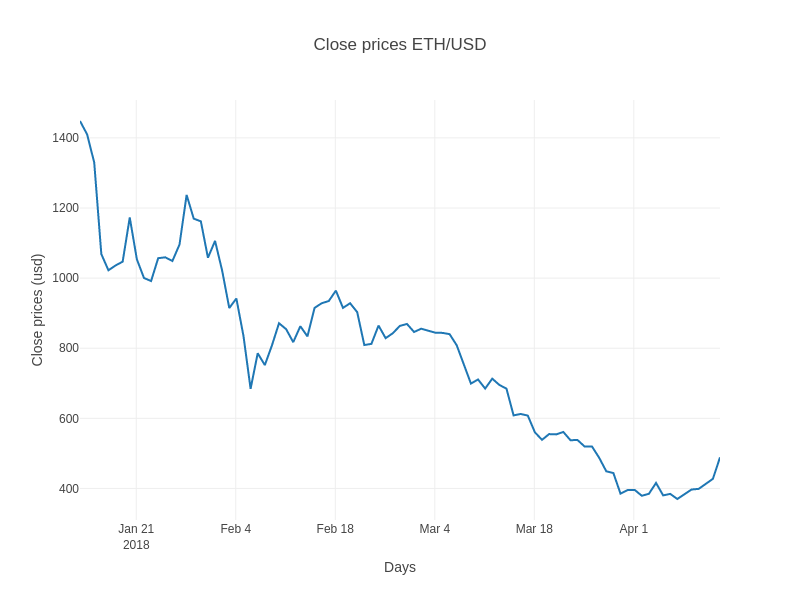}
    \caption{{\scriptsize 13/01/2018-14/04/2018}}
    \label{fig:eth-cp}
  \end{subfigure}
%  \medskip
  \begin{subfigure}{0.6\textwidth}
    \centering
    \includegraphics[width=\textwidth]{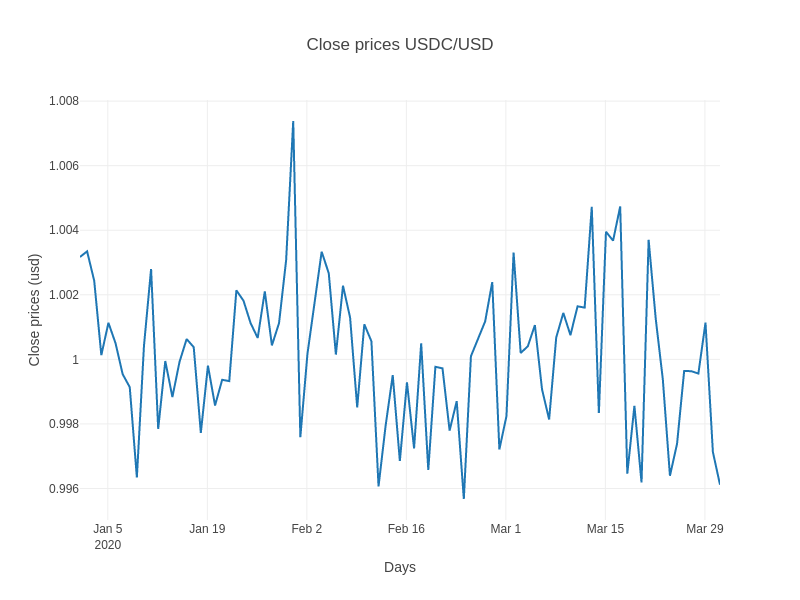}
    \caption{{\scriptsize 01/01/2020-01/04/2020}}
    \label{fig:usdc-cp}
  \end{subfigure}
%  \medskip
  % \end{figure}
  % \begin{figure} \ContinuedFloat
  %   \centering
  \begin{subfigure}{0.6\textwidth}
    \centering
    \includegraphics[width=\textwidth]{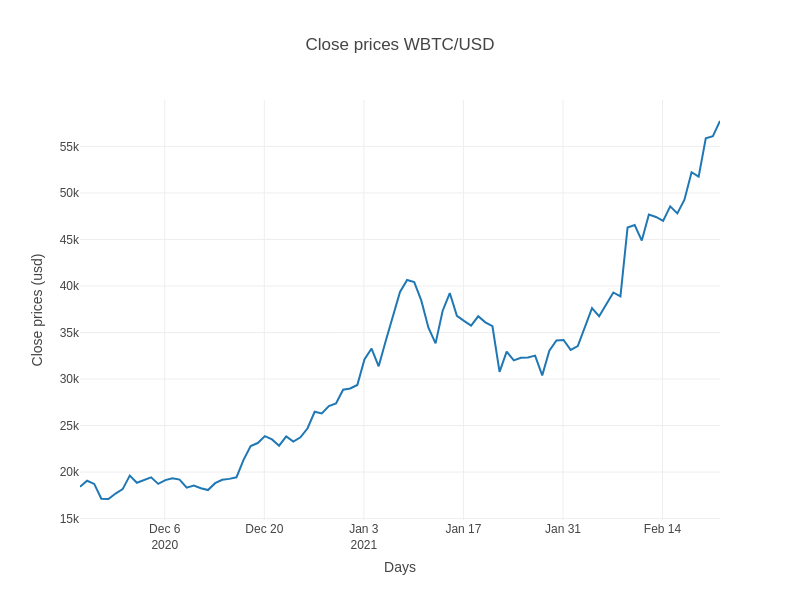}
    \caption{{\scriptsize 24/11/2020-23/02/2021}}
    \label{fig:wbtc-cp}
  \end{subfigure}
  \caption{Trimester closing prices, collected from \href{https://www.coingecko.com/api/documentations/v3}{CoinGecko APIs}
  %\comav{Plots way too small. I made them bigger (from .32 to .495). But the font is still too small: the python script should be changed with larger fonts.}
  }
  \label{fig:hist-cp}
\end{figure}

\clearpage

\begin{figure}
  \centering
  \begin{subfigure}{0.6\textwidth}
    \centering
    \includegraphics[width=\textwidth]{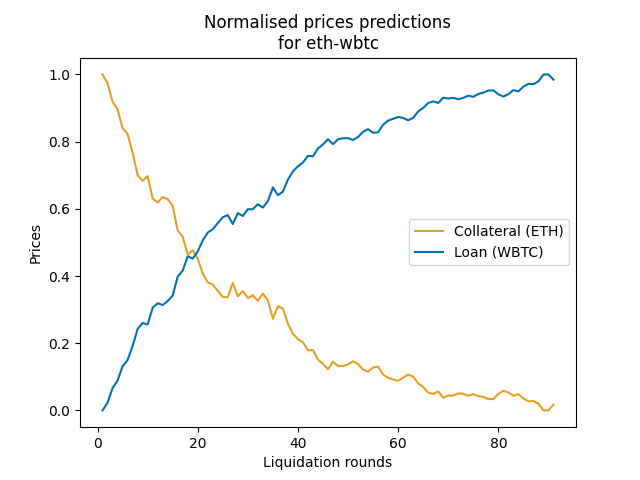}
    \caption{}    
    % \caption{Predictions based on GBMs instantiated respectively with parameters ETH for the collateral asset and WBTC for the loan asset.}
    \label{fig:norm-eth-wbtc}
  \end{subfigure}
%  \medskip       
%  
  \begin{subfigure}{0.6\textwidth}
    \centering
    \includegraphics[width=\textwidth]{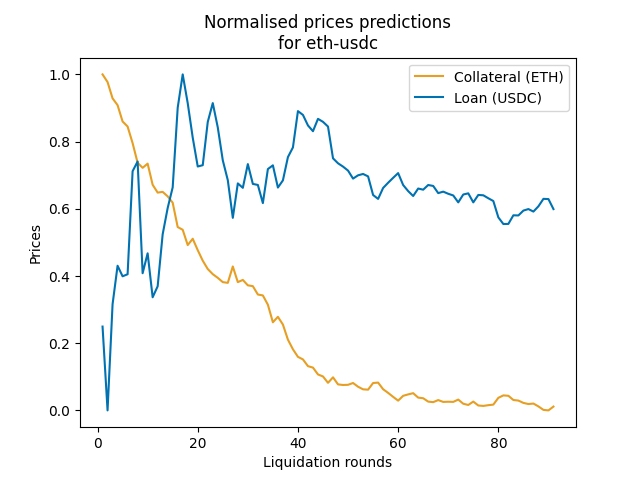}
    \caption{}
    % \caption{Predictions based on GBMs instantiated respectively with parameters ETH for the collateral asset and USDC for the loan asset.}
    \label{fig:norm-eth-usdc}
  \end{subfigure}
%  \medskip
  % \end{figure}
  % \begin{figure}[h] \ContinuedFloat
  %   \centering
  \begin{subfigure}{0.6\textwidth}
    \centering
    \includegraphics[width=\textwidth]{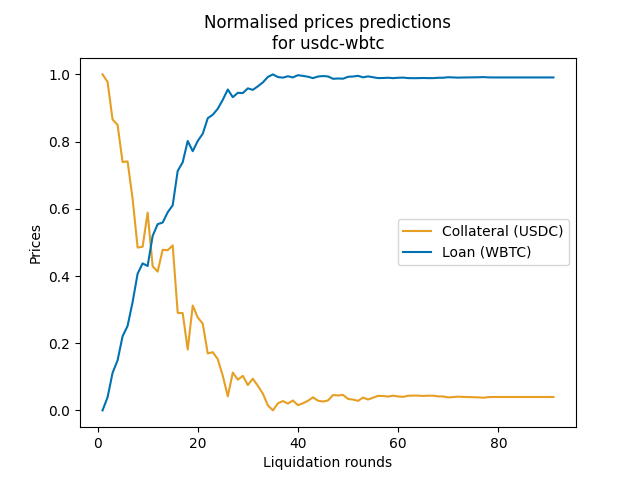}
    \caption{}
    % \caption{Predictions based on GBMs instantiated respectively with parameters USDC for the collateral asset and WBTC for the loan asset.}         
    \label{fig:norm-usdc-wbtc}
  \end{subfigure}
  \caption{Prices predictions produced, for each scenario in \Cref{tab:prices-pairs}, by GBMs instantiated with the parameters in \Cref{tab:gbm-params}.
  %\comav{Plots were way too small. I made them bigger (from .32 to .495). But the person who created them should modify the python script to increase the fonts}
  }
  \label{fig:norm-all}
\end{figure}

\clearpage

\begin{figure}
  \centering
  \begin{subfigure}[H]{0.75\textwidth}
    \centering
    \adjincludegraphics[width=1.0\textwidth,trim={0 {0.19\height} 0 {0.22\height}},clip]{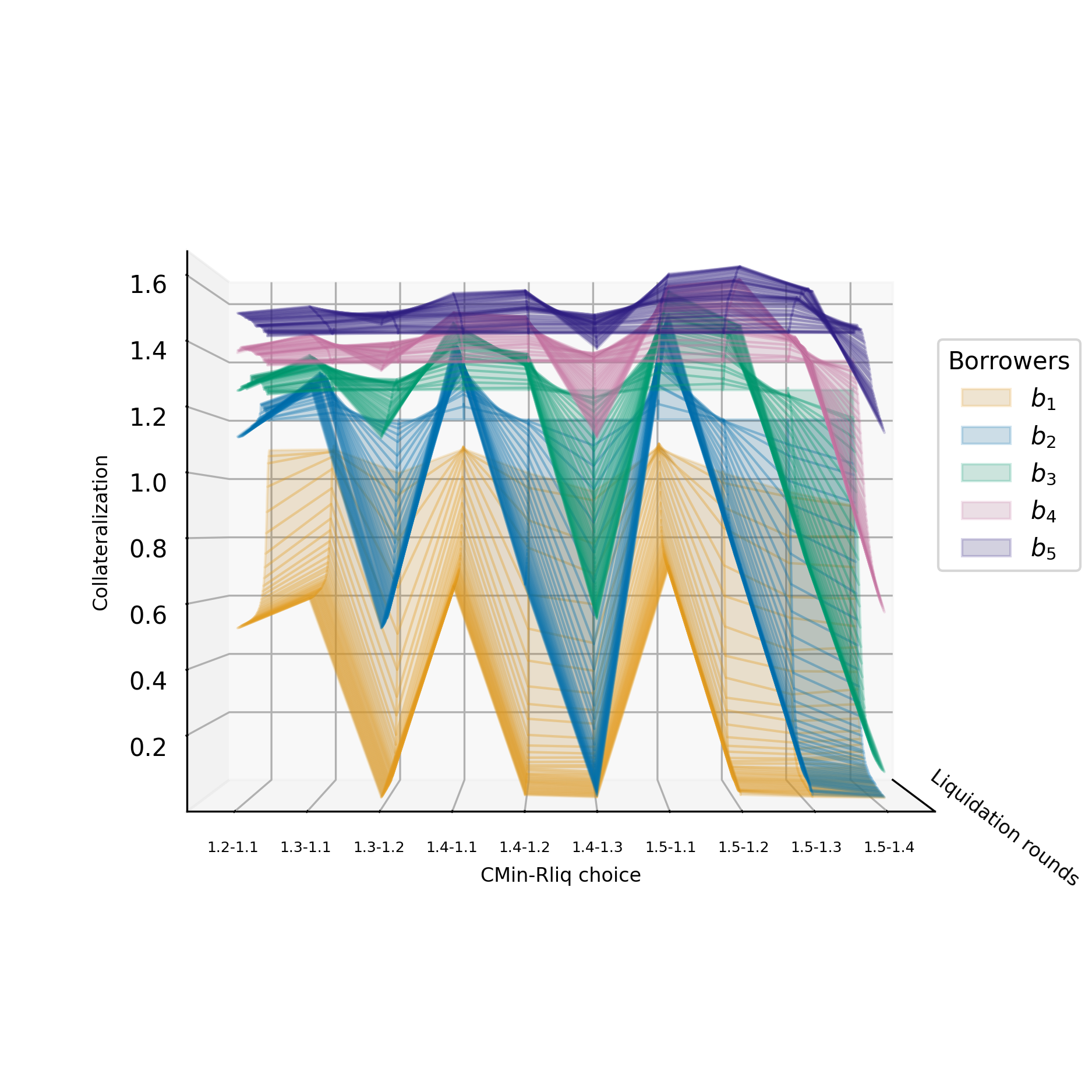}
    \caption{Scenario: eth-wbtc.}
    \label{fig:s-eth-wbtc-cmin-rliq-ag_1_5}
  \end{subfigure}
% \end{figure}
% \begin{figure} \ContinuedFloat
%   \centering
  \medskip
  \begin{subfigure}[H]{0.75\textwidth}
    \centering
    \adjincludegraphics[width=1.0\textwidth,trim={0 {0.19\height} 0 {0.22\height}},clip]{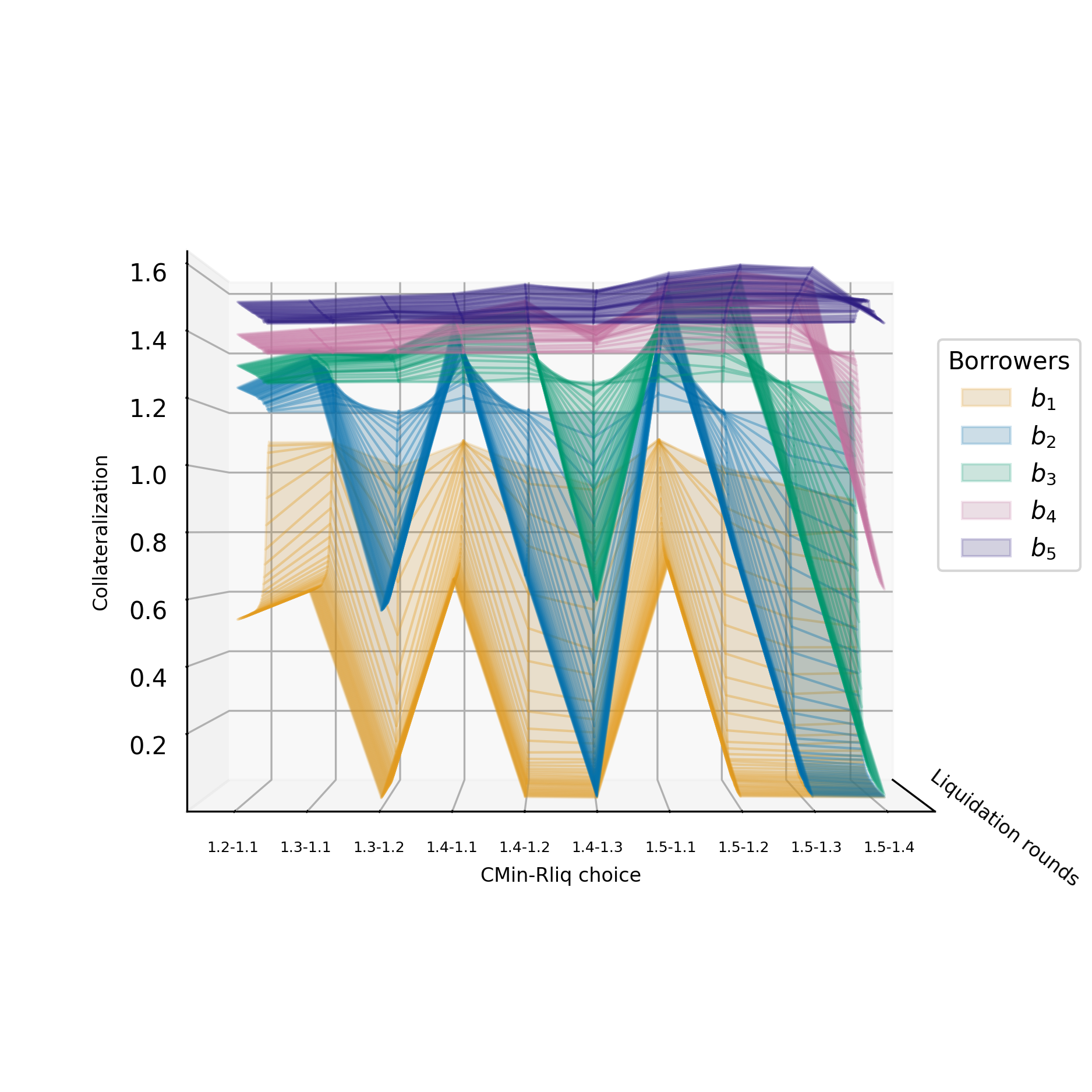}
    \caption{Scenario: eth-usdc.}
    \label{fig:s-eth-usdc-cmin-rliq-ag_1_5}
  \end{subfigure}
  \medskip 
% \end{figure}
% \begin{figure} \ContinuedFloat
  \begin{subfigure}[H]{0.75\textwidth}
    \centering
    \adjincludegraphics[width=1.0\textwidth,trim={0 {0.19\height} 0 {0.22\height}},clip]{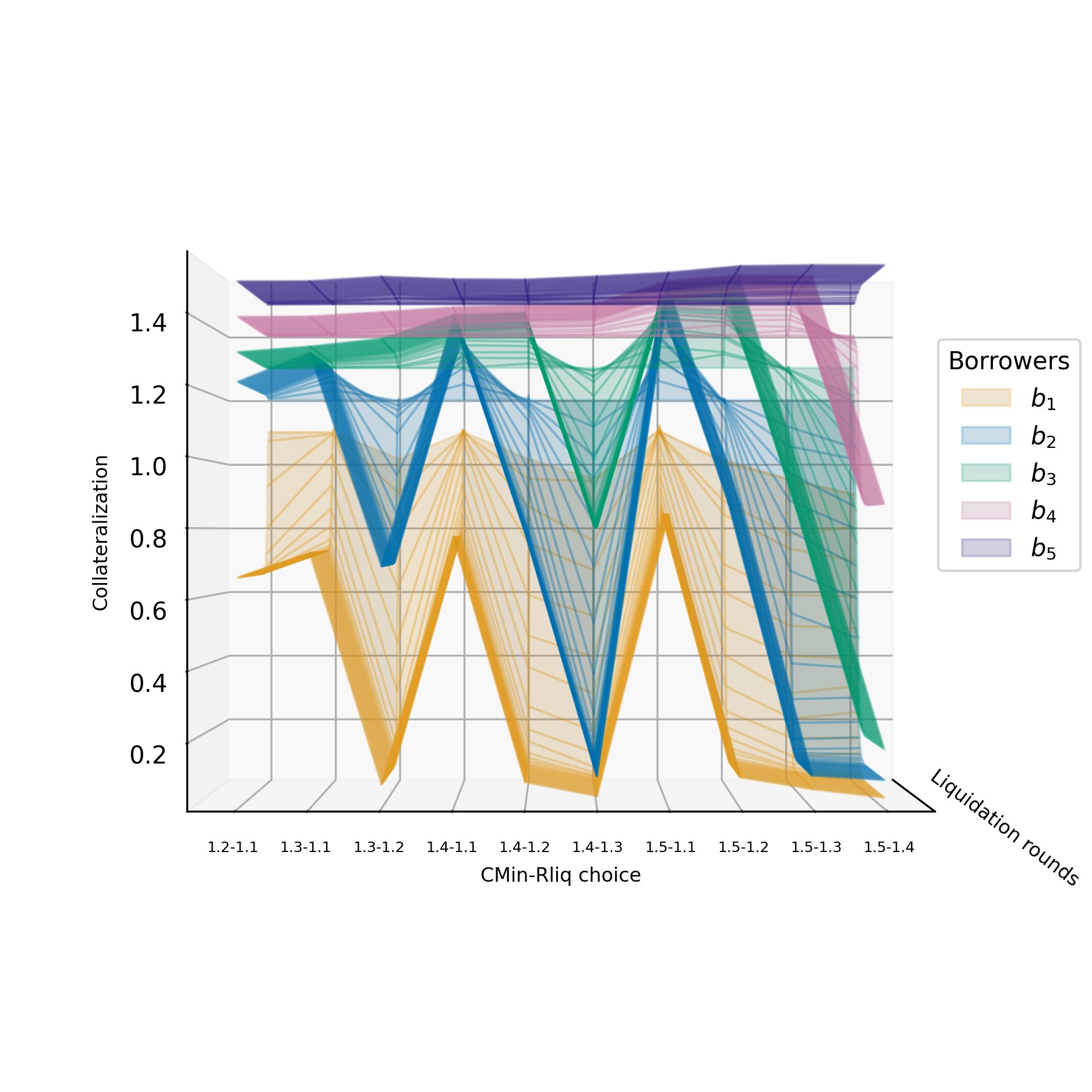}
    \caption{Scenario: usdc-wbtc.}
    \label{fig:s-usdc-wbtc-cmin-rliq-ag_1_5}
  \end{subfigure}
  \caption{Per-borrower collateralization ($\mathrm{b_{1}}$ to $\mathrm{b_{5}}$) in the three prices scenarios, for varying \texttt{CMin-rliq} choices.}
  \label{fig:3d-1-5}
\end{figure}

\clearpage

\begin{figure}
  \centering
  \begin{subfigure}[H]{0.75\textwidth}
    \centering
    \adjincludegraphics[width=1.0\textwidth,trim={0 {0.19\height} 0 {0.22\height}},clip]{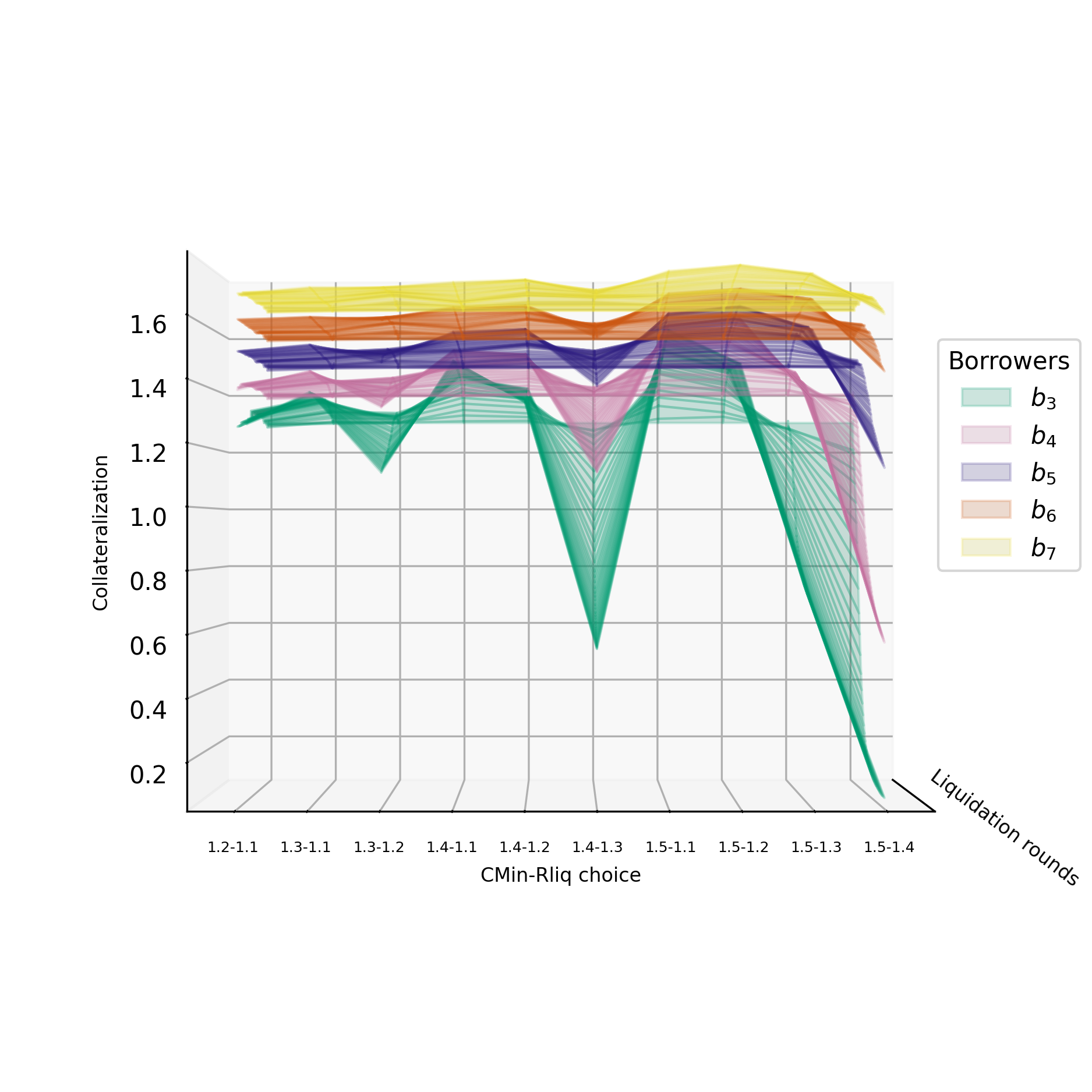}
    \caption{Scenario: eth-wbtc.}
    \label{fig:s-eth-wbtc-cmin-rliq-ag_3_7}
  \end{subfigure}
% \end{figure}
% \begin{figure} \ContinuedFloat
%   \centering
  \medskip
  \begin{subfigure}[H]{0.75\textwidth}
    \centering
    \adjincludegraphics[width=1.0\textwidth,trim={0 {0.19\height} 0 {0.22\height}},clip]{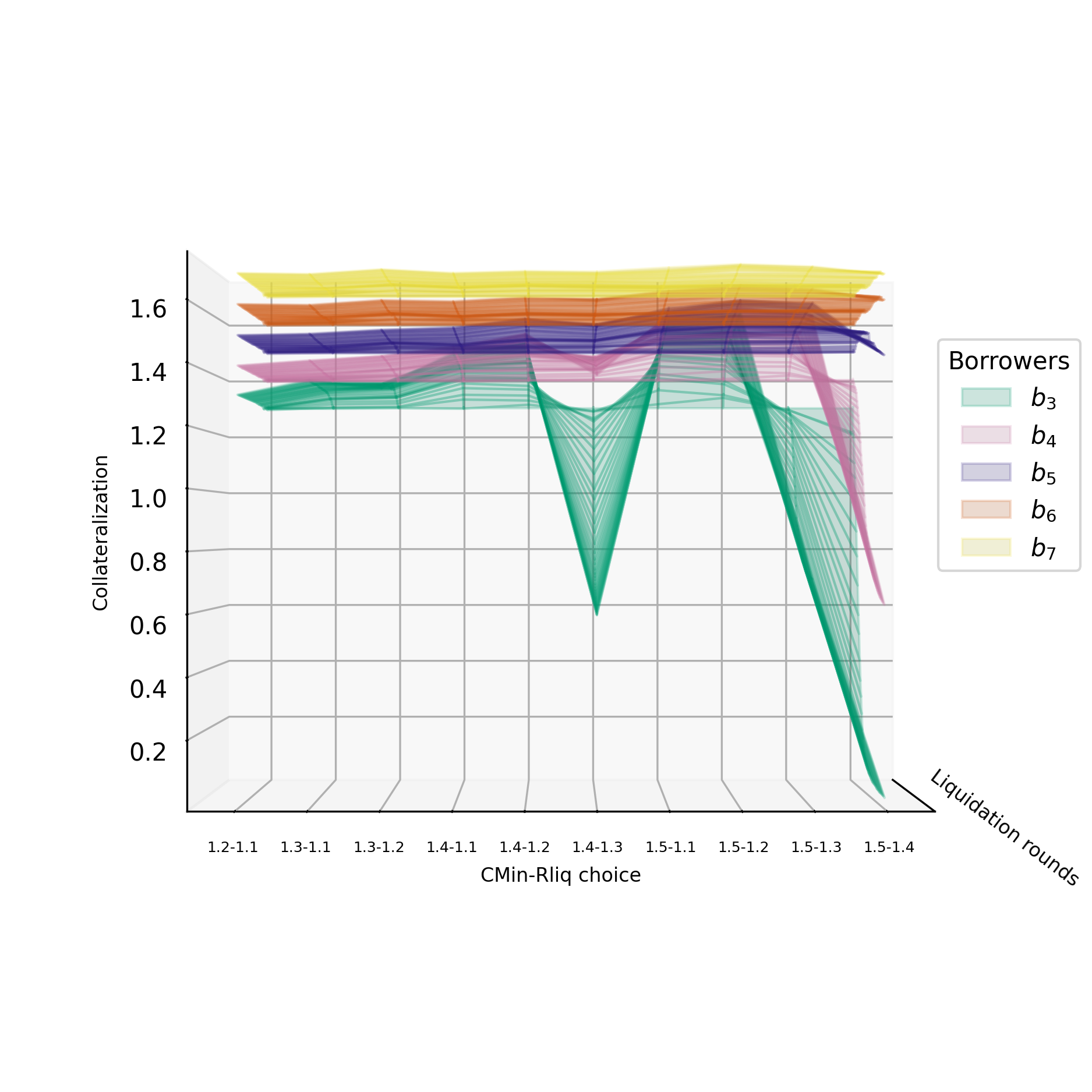}
    \caption{Scenario: eth-usdc.}
    \label{fig:s-eth-usdc-cmin-rliq-ag_3_7}
  \end{subfigure}
  \medskip 
% \end{figure}
% \begin{figure} \ContinuedFloat
  \begin{subfigure}[H]{0.75\textwidth}
    \centering
    \adjincludegraphics[width=1.0\textwidth,trim={0 {0.19\height} 0 {0.22\height}},clip]{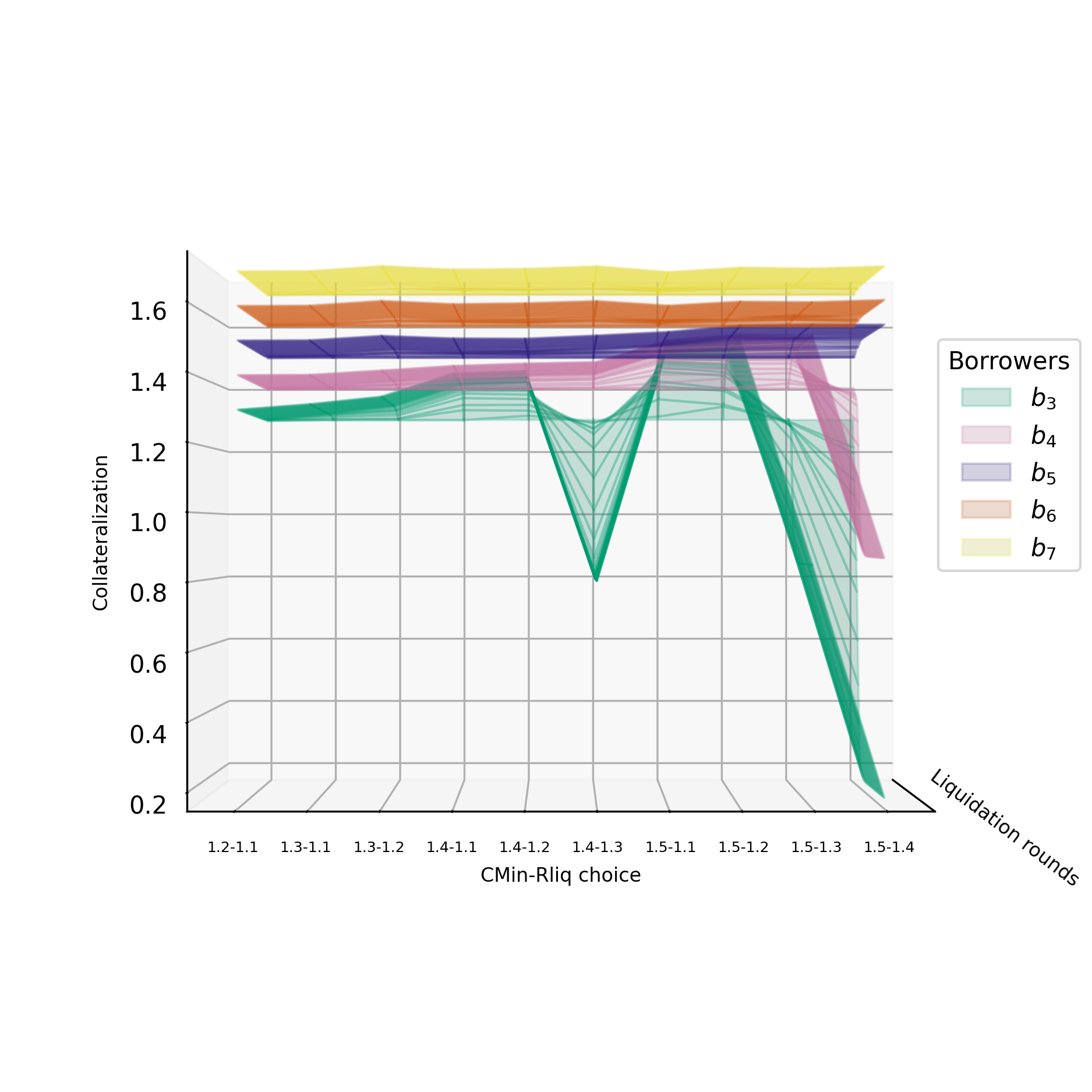}
    \caption{Scenario: usdc-wbtc.}
    \label{fig:s-usdc-wbtc-cmin-rliq-ag_3_7}
  \end{subfigure}
  \caption{Per-borrower collateralization ($\mathrm{b_{3}}$ to $\mathrm{b_{7}}$) in the three prices scenarios, for varying \texttt{CMin-rliq} choices.
  %\comav{Plots are fine, but the font should be much bigger: we need to modify the python scripts}
  }
  \label{fig:3d-3-7}
\end{figure}

\end{document}